\documentclass[useAMS,usenatbib]{mn2e}
\usepackage{graphicx}
\usepackage{color}
\usepackage{ulem}
\usepackage{pdflscape}
\def\apj{ApJ}
\def\apjs{ApJS}
\def\apjl{ApJL}
\def\aap{A\&A}

\def\assp{Astrophysics and Space Science Proceedings}
\def\aj{AJ}
\def\mnras{MNRAS}

\def\pasa{PASA}
\def\pasj{PASJ}
\def\pasp{PASP}
\def\nat{Nature}

\def\araa{Ann.Rev.Astron.Astrophys.}

\def\rmxaa{Revista Mexicana de Astronomia y Astrofisica}

\title[VLT/XShooter spectra of five low-z LyC leakers]
{Properties of five z$\sim$0.3--0.4 confirmed LyC leakers: VLT/XShooter
observations}        
\author[N. G. Guseva et al.]{N. G.\ Guseva$^{1}$\thanks{E-mail: nguseva@bitp.kiev.ua}, Y. I.\ Izotov$^{1}$,
D.\ Schaerer$^{2,3}$, J. M.\ V\'ilchez$^{4}$, R.\ Amor\'in$^{5,6}$,
\newauthor
E.\  P\'erez-Montero$^{4}$, J.\ Iglesias-P\'aramo$^{4,7}$, A.\ Verhamme$^{2}$, C.\ Kehrig$^{4}$,
\newauthor
L.\ Ramambason$^{8}$\\
                $^{1}$Bogolyubov Institute for Theoretical Physics,
                     Ukrainian National Academy of Sciences,
                     Metrologichna 14b, Kyiv 03143,  Ukraine,\\
                $^{2}$Observatoire de Gen\`eve, Universit\'e de Gen\`eve, 
                     51 Ch. des Maillettes, 1290, Versoix, Switzerland,\\
                $^{3}$IRAP/CNRS, 14, Av. E. Belin, 31400 Toulouse, France\\
                $^{4}$Instituto de Astrof\'isica de Andaluc\'ia - CSIC, Glorieta
                     de la Astronom\'ia s/n, 18008 Granada, Spain\\
                $^{5}$Instituto de Investigaci\'on Multidisciplinar en Ciencia y
                     Tecnolog\'ia, Universidad de La Serena, Ra\'ul Bitr\'an
                     1305, La Serena, Chile\\
                $^{6}$Departamento de Astronom\'ia, Universidad de La Serena,
                     Av. Juan Cisternas 1200 Norte, La Serena, Chile\\
                $^{7}$Estaci\'on Experimental de Zonas \'Aridas - CSIC, Ctra. de
                     Sacramento s/n, 04120 Almer\'ia, Spain \\
                $^{8}$Universit\'e Paris-Saclay, CEA, D\'epartement
                     d'Astrophysique, 91191, Gif-sur-Yvette, France\\
}
\begin{document}

\pagerange{\pageref{firstpage}--\pageref{lastpage}} \pubyear{2020}

\maketitle

\label{firstpage}

\begin{abstract}
  Using new VLT/XShooter spectral observations we analyse the physical
properties 
of five $z$~$\sim$~0.3~-~0.4 confirmed LyC leakers.
  Strong resonant Mg~{\sc ii}~$\lambda$$\lambda$2796,2803\AA\ emission lines
 ($I(\lambda$$\lambda$2796,2803)/$I$(H$\beta$) $\simeq$ 10-38 per cent) and 
non-resonant Fe~{\sc ii}$^*$~$\lambda$$\lambda$2612,2626\AA\ emission lines
are observed in spectra of five and three galaxies, respectively.
We find high electron densities $N_e \sim 400$ cm$^{-3}$, significantly higher than in typical
low-$z$, but comparable to those measured in $z \sim 2-3$ star-forming galaxies.
  The galaxies have a mean value of log N/O = --1.16, close to the maximum values 
  found for star-forming (SF) galaxies in the metallicity range of 
12~+~log~O/H $\simeq$ 7.7~-~8.1.
  All 11 low-$z$ LyC emitting galaxies found by Izotov et al. (2016, 2018),
including the ones considered in the present study,
are characterised by high EW(H$\beta$)$\sim$200-400\AA, high
ionisation parameter (log($U$) = --2.5 to --1.7), high average ionising
photon production efficiency $\xi$ = 10$^{25.54}$ Hz erg$^{-1}$
and hard ionising radiation.
  On  the BPT diagram we find the same offset of our leakers 
from low-$z$ main-sequence SFGs as that for local analogues of LBGs
and extreme SF galaxies at $z$~$\sim$~2~-~3.
  We confirm the effectiveness of the He~{\sc i} emission lines diagnostics 
proposed by Izotov et al. (2017) in searching for LyC leaker candidates 
and find that their intensity ratios correspond to those in a median with 
low neutral hydrogen column density
$N$(H~{\sc i}) = 10$^{17}$ - 5$\times$10$^{17}$ cm$^{-2}$ that permit leakage 
of LyC radiation, likely due to their density-bounded H {\sc ii} regions.
\end{abstract}

\begin{keywords}
galaxies: dwarf -- galaxies: starburst -- galaxies: ISM -- galaxies: abundances.
\end{keywords}

\section{Introduction}\label{sec:Introduction}

  Recently \citet{I16a,I16b,I18a,Izotov2018b}
have discovered significant emission of Lyman continuum (LyC) ionising 
radiation leaking with the escape fractions of 2-76 per cent
in a sample of 11 low-$z$ compact active 
star-forming galaxies (SFGs) observed with the {\sl Hubble Space Telescope} 
({\sl HST}) in conjunction with the Cosmic Origins Spectrograph (COS).
These galaxies hereafter refereed to as LyC leakers, possess many properties similar
to those of high-redshift galaxies both at $z$ $\sim$ 2-3 and 
$z$ $\ga$ 6 such as compact morphology with similar galaxy radii
\citep*[e.g. ][] {Bouwens2004,Ferguson2004,Oesch2010,Ono2012,Shibuya2015,CurtisLake2016,Paulino-Afonso2018},
strong emission lines with 
high EWs
\citep*[e.g. ][]{SchaererdeBarros2009,Smit2014,Smit2015,Roberts-Borsani2016,Bowler2017,Castellano2017,Fletcher2019,Bian2020a,Endsley2020},
  similar low stellar masses, low metallicities and high 
  specific star formation rates
\citep*[e.g. ][]{Stark2013a,Nakajima2013,JaskotOey2013,deBarros2014,Gonzalez2014,Duncan2014,Nakajima2014,Becker2015,Grazian2015,Salmon2015,Huang2016,Stark2016,Stark2017,Santini2017,Dors2018},
    small dust content
\citep*[e.g. ][]{Ouchi2013,Ota2014,Schaerer2015,Maiolino2015,Watson2015}    
 and are considered as the main sources of reionisation
of the Universe after the cosmic ``Dark Ages''. This makes low-$z$ LyC leakers
the best local analogues of reionisation galaxies
\citep*[see e.g. ][]{Stark2016,Schaerer2016,Ma2020}.
Given their proximity, these galaxies represent excellent laboratories
for a detailed study of their physical conditions, and the main 
mechanisms responsible for 
LyC leakage. Ground-based
spectroscopic observations in the visible and near-infrared ranges are 
necessary for that.

\begin{table*} 
 \caption{General characteristics of our sample galaxies \label{tab3}}
 \begin{tabular}{lccccccccccc} \hline \hline
  {\sc Galaxy}&
  \multicolumn{1}{c}{R.A.(J2000)} & \multicolumn{1}{c}{Dec.(J2000)} &
  \multicolumn{1}{c}{z} 
  &\multicolumn{1}{c}{$f_{\rm esc}$(LyC)$^{\rm a}$}
  &\multicolumn{1}{c}{log $t$$^{\rm b}$}
  &\multicolumn{1}{c}{log $M_{\star}$$^{\rm c}$}
  &\multicolumn{1}{c}{SFR$^{\rm d}$}
  &\multicolumn{1}{c}{EW(H$\beta$)$^{\rm e}$}
  &\multicolumn{1}{c}{12+logO/H$^{\rm f}$}
  &\multicolumn{1}{c}{$^{\rm g}$} \\ \hline
 J0901$+$2119 & 09:01:45.61&$+$21:19:27.78&0.2993&  2.7 & 2.4 &  9.80 & 20 & 357.5 & 8.05 & Fe {\sc ii}$^*$  \\
 J0925$+$1403 & 09:25:32.37&$+$14:03:13.06&0.3010&  7.8 & 2.6 &  8.91 & 52 & 208.1 & 8.12 & Fe {\sc ii}$^*$  \\ 
 J1011$+$1947 & 10:11:38:28&$+$19:47:20.94&0.3322& 11.4 & 3.4 &  9.00 & 24 & 314.9 & 7.97 & ... \\
 J1154$+$2443 & 11:54:48.85&$+$24:43:33.03&0.3690& 46.0 & 2.6 &  8.20 & 19 & 255.1 & 7.75 & ... \\
 J1442$-$0209 & 14:42:31.39&$-$02:09:52.03&0.2937&  7.3 & 3.4 &  8.96 & 36 & 186.8 & 7.98 & Fe {\sc ii}$^*$  \\ 
 \hline
 
\end{tabular}
 
 \hbox{$^{\rm a,b,c,d}$Data obtained by \citet{I16a,I16b,I18a,Izotov2018b}.}

 \hbox{$^{\rm a}$Lyman continuum escape fraction in per cent.}

\hbox{$^{\rm b}$Starburst (SB) age in Myrs. Ages and stellar masses are derived from SED fitting of the SDSS spectra.}

\hbox{$^{\rm c}$$M_{\star}$ is the galaxy stellar mass in units of M$_\odot$.}

\hbox{$^{\rm d}$Star formation rate obtained from the H$\beta$ luminosity according to \citet{Kennicutt1998} in units of M$_\odot$ yr$^{-1}$.}

\hbox{$^{\rm e,f}$Data obtained in this paper.}

\hbox{$^{\rm e}$Equivalent width of H$\beta$ in \AA.}

\hbox{$^{\rm g}$Detection of Fe {\sc ii}$^*$ emission.}

  \end{table*}

   In this paper we present a spectroscopic study with the European Southern
Observatory (ESO) Very Large Telescope (VLT) in conjunction with the XShooter 
spectrograph of five LyC leaking galaxies from the 
\citet{I16a,I16b,I18a,Izotov2018b} sample with LyC 
escape fractions $f_{\rm esc}$(LyC) = 2.7 - 46.0 per cent and low enough 
declinations
($<$ $+$25~deg), making them accessible for observations at Paranal.
Throughout of the text $f_{\rm esc}$(LyC) is the absolute escape fraction,
which is defined as a ratio of the observed LyC flux corrected
for Milky Way (MW) extinction, and the intrinsic galaxy LyC flux.
  Such observations provide an excellent opportunity for a comprehensive study
of  selected galaxies over a wide wavelength range ($\sim$ $\lambda$3000 - 24000\AA).  
  The basic properties of our galaxy sample are summarised in Table~\ref{tab3}.

\begin{table*}
  \caption{Log of observations \label{tab6}}
  \begin{tabular}{lccccccl}  \hline \hline
  {Name}&
\multicolumn{1}{c}{Date} &
\multicolumn{3}{c}{Exposure time$^{\rm a}$} & \multicolumn{1}{c}{Airmass$^{\rm b}$} &
\multicolumn{1}{c}{Seeing$^{\rm c}$}&\multicolumn{1}{c}{Spectrophotometric standard stars$^{\rm d}$}\\
 &&UVB&VIS&NIR&& \\ \hline

J0901$+$2119 &2019-01-09&8280&8640&1500&1.47&0.77&GD71(1.69) \\
J0925$+$1403 &2019-01-15&2760&2880& 496&1.45&0.89&GD71(1.69) \\
J1011$+$1947 &2019-01-09&5520&5760&1000&1.40&0.93&GD71(1.69); LTT3218(1.36) \\
J1154$+$2443 &2019-04-30&8280&8640&1500&1.54&1.10&EG274(1.34); LTT3218(1.38); LTT7987(1.02) \\
J1442$-$0209 &2019-05-24&2760&2880& 500&1.08&0.99&LTT3218(1.08) \\

\hline  \hline
  \end{tabular}

  \hbox{$^{\rm a}$In sec.}  
  \hbox{$^{\rm b}$Average airmass during observation.} 
  \hbox{$^{\rm c}$Average seeing (FWHM) in arcsec.}
  \hbox{$^{\rm d}$Spectrophotometric standard stars used for the flux 
calibration. They were observed at average airmasses shown in parentheses.}

\end{table*}

   This paper is organised as follows.
   In Section~\ref{sec:Reduction} we describe the VLT/XShooter
spectrophotometric observations and 
data reductions. 
   In Section~\ref{sec:Results} we present the results obtained with the new
observations.
  The element abundance determination with emphasizing 
the problems of nitrogen abundance is given in
Subsection~\ref{subsec:abundances}.
   High intensities of nebular helium emission line
He {\sc ii}~$\lambda$4686\AA\  are considered in Subsection \ref{subsec:HeII}.
   In Subsection \ref{subsec:Mg} we discuss resonant
Mg~{\sc ii}~$\lambda$$\lambda$2796,2803\AA\ emission lines.
   The position of confirmed local LyC leakers on the BPT diagram and their
ionisation parameters and ionising photon production efficiencies are discussed in
Subsection \ref{subsec:BPT}.   
   In Subsection \ref{subsec:HeIdiagnostic} we present the He~{\sc i} emission 
line diagnostics for the LyC leakers. 
   Finally, in Section \ref{subsec:summary} we summarise our main results.

\section{Observations and data reduction}\label{sec:Reduction}

   Spectral observations of five confirmed LyC leakers were carried out with the
XShooter spectrograph mounted at the UT2 Cassegrain focus of the
VLT in nodding-on-slit mode 
during 2019 (ESO Program ID 0102.B-0942(A)).
   The use of three UVB (1 arcsec $\times$ 11 arcsec slit, 
$R$ $\approx$ 5100), VIS (0.9 arcsec $\times$ 11 arcsec slit, 
$R$ $\approx$ 8800) and NIR (0.9 arcsec $\times$ 11 arcsec slit, 
$R$ $\approx$ 5100) arms made it possible to obtain the
spectrum of each object simultaneously over a wide wavelength range,
in particular in the UVB, VIS and NIR arms with wavelength ranges
$\sim$$\lambda$3000-5600\AA, $\sim$$\lambda$5500-10200\AA\ and
$\sim$$\lambda$10200-24000\AA, respectively.
All observations were obtained during clear nights.
We note, that two galaxies J1011$+$1947 and J1154$+$2443 were observed
during several nights. For the flux calibration, several standard stars
(GD71, LTT3218, EG274, LTT7987) were observed at various airmasses in the
range 1.02 - 1.69.
The log of observations is presented in Table~\ref{tab6}.

  {\sc iraf}\footnote{{\sc iraf} 
is distributed by the 
National Optical Astronomy Observatories, which are operated by the Association
of Universities for Research in Astronomy, Inc., under cooperative agreement 
with the National Science Foundation.} was used to reduce the observations and,
as the first step,  to subtract the
bias in the UVB and VIS arms and dark frames in the NIR arm.
   Applying the {\it crmedian}   routine to all UVB, VIS and NIR arms we removed
the cosmic rays.
  We applied the correction for telluric absorption of the galaxy spectra
in the wavelength ranges which include [S~{\sc ii}]$\lambda$6717, 6731
emission lines. To do this we produce the normalised spectrum of the standard
star from its observed spectrum, i.e. adopting its continuum equal to unity.
Then the spectrum of the galaxy corrected for telluric absorption is derived
dividing the observed galaxy spectrum by the normalised standard star spectrum.
However, no correction has been done for [S~{\sc iii}] $\lambda$9069, 9531
emission lines because of stronger and variable telluric bands of H$_2$O. 
   The correction for the pixel sensitivity,
background subtraction, wavelength calibration, correction for distortion and
tilt of each frame were performed.
    After this, the one-dimensional spectra were extracted from the
two-dimensional frames in apertures of 1.6 arcsec along the slit for UVB and VIS
arms and 2.4 arcsec for NIR arm.
    The flux-calibrated rest-frame spectra are shown in Fig.~\ref{fig1}.

  Emission line fluxes and their errors were measured in
flux-calibrated and non-flux-calibrated spectra, respectively,
using 
total integral fluxes with the {\sc iraf} {\it splot}  routine
\citep*[see for more details, e.g. ][]{Guseva2012,IzTGuseva2014,Guseva2015}. 
     The internal extinction and underlying hydrogen stellar absorption were
derived 
iteratively from the Balmer decrement following \citet*{Izotov1994} and adopting
the \citet*{Cardelli1989} reddening law with $R$($V$) = 3.1, 
after correcting the spectra for  Milky Way extinction.
  The MW extinction correction was applied to the spectrum at observed
  wavelengths adopting the extinction $A$($V$) from the NASA/IPAC Extragalactic
  Database (NED) and the same value of 
$R$($V$). It was also assumed that EWs of absorption lines are the same 
for all hydrogen Balmer transitions.
  The equivalent widths of the Balmer absorption lines range from 2.5
to 3.5\AA,  compatible with the predictions of the evolutionary
stellar population synthesis models by \citet{Gonzalez1999} for young
starbursts. 

   Extinction-corrected fluxes $I$($\lambda$) relative to H$\beta$
multipled by 100, equivalent widths of emission lines,  equivalent width of
underlying hydrogen absorption lines EW(abs), the extinction coefficient
$C$(H$\beta$) and the observed flux of H$\beta$ for each galaxy are given in 
Table~\ref{tab1}. 

\section{Results}\label{sec:Results}

\subsection{Physical properties and element abundances}\label{subsec:abundances}

  To derive physical conditions and element abundances we follow prescriptions
by \citet{IzStasMeynet2006a} \citep*[see also e.g. ][]{IzTh2004,Izotov2019}.
Briefly, \citet{IzStasMeynet2006a} adopt the three zone model of the
H~{\sc ii} region with respective electron temperatures $T_{\rm e}$(O~{\sc iii}),
$T_{\rm e}$(S~{\sc iii}) and $T_{\rm e}$(O~{\sc ii}). The electron temperature
$T_{\rm e}$(O~{\sc iii}) is derived from the ratio of 
[O~{\sc iii}] line fluxes $\lambda$4363/$\lambda$(4959+5007) in the
high-ionisation zone. It is used to obtain abundances of ions O$^{2+}$, 
Ne$^{2+}$ and Ar$^{3+}$.
   Electron temperatures $T_{\rm e}$(O~{\sc ii}) and $T_{\rm e}$(S~{\sc iii})
are derived from relations obtained from photoionisation models of
H~{\sc ii}  regions.
  The electron temperature $T_{\rm e}$(S~{\sc iii}) in the
intermediate-ionisation zone is used to derive the abundances of ions S$^{2+}$,
and Ar$^{2+}$. 
   The electron temperature $T_{\rm e}$(O~{\sc ii}) in the
low-ionisation zone is used to derive the
abundances of ions O$^+$, N$^+$, S$^+$, Mg$^+$ and Fe$^{2+}$.
     The sulfur emission line ratio [S~{\sc ii}]$\lambda$6717/$\lambda$6731
is used to derive the electron number density $N_{\rm e}$(S {\sc ii}).
  The total heavy element abundances are obtained with the
use of ionisation correction factors ({\sl ICF}s) by
 \citet{IzStasMeynet2006a}.

   For comparison we also provide the electron temperatures and electron number
densities obtained with {\sc iraf} routine \textit{temden}. 
  Additionally we use Eq.7 by \citet{Sanders2016} for electron number density determinations
from the [O~{\sc ii}]$\lambda$3729/$\lambda$3726 and
[S~{\sc ii}]$\lambda$6717/$\lambda$6731  line ratios with new collision strengths
of \citet{Tayal2007} for [O~{\sc ii}] and \citet{TayalZat2010} for [S~{\sc ii}]
and with transition probabilities for both species of \citet{Fische2014}. 
It is worth emphasizing that even small changes in flux ratios of components of
[O~{\sc ii}] doublet,
as it is in the case of {\sc iraf} \textit{temden} or method by 
\citet{Sanders2016} can result in large errors of the electron number density.

  In the XShooter observations, the two components of [O~{\sc ii}] are resolved but 
slightly blended, and the [S~{\sc ii}] doublet often falls into the wavelength
region of strong telluric absorption lines.
   For example, in the case of J1442$-$0209 the
[S~{\sc ii}] is shifted to the region of night sky O$_2$ band at rest-frame
wavelengths $\lambda$$\lambda$8700-8715\AA.
   [S~{\sc ii}] lines in spectra of J1011+1947 and J1154+2443 are in the forest 
of telluric absorption lines as it is seen in the spectra of a standard star.
  $T_{\rm e}$(S~{\sc iii}) cannot be determined from the XShooter spectra using
the [S~{\sc iii}]$\lambda$9069 and $\lambda$9531 lines, since
these lines fall into the wavelength region of strong telluric absorption.
   The correction for the telluric absorption introduces uncertainties 
influencing the number density values in
Table~\ref{tab2}.
  Based on this, we calculated the ionic and element abundances using the
 method  
 by \citet{IzStasMeynet2006a}, the values of $T_{\rm e}$(O~{\sc iii}),
$T_{\rm e}$(O~{\sc ii}), $T_{\rm e}$(S~{\sc iii}) in the first three lines of
Table~\ref{tab2} and $N_{\rm e}$(S~{\sc ii}) in the sixth line of the same 
Table 
 with the exception of galaxies
J1154+2443 and J1442$-$0209, for which $N_e$ were taken from the
[O~{\sc ii}]$\lambda$3726,3729 doublet ratio adopting the average of 
highlighted values in Table~\ref{tab2}). 
   Electron temperatures $T_{\rm e}$(O~{\sc iii}), $T_{\rm e}$(O~{\sc ii}),
$T_{\rm e}$(S~{\sc iii}), electron number densities $N_{\rm e}$(S~{\sc ii})
and $N_{\rm e}$(O~{\sc ii}), 
ionic abundances, ionisation correction factors {\sl ICF}s and element
abundances of oxygen, nitrogen, neon, sulfur, argon, iron and
magnesium are given in Table~\ref{tab2}. 

    Using most reliable determinations of $N_e$ in high-$z$ SFGs from
\citet{Christensen2012,Stark2013,James2014,Bayliss2014,Steidel2014,Steidel2016,Sanders2020} with 
element abundances derived by the direct $T_e$ method 
and dividing them into groups by distance/redshift we obtain average electron number
densities
$N_e$ $\sim$ 260 cm$^{-3}$ ($z$ $\sim$ 1.4), $N_e$ $\sim$ 460 cm$^{-3}$
($z$ $\sim$ 2.3) and $N_e$ $\sim$ 450 cm$^{-3}$ ($z$ $\sim$ 3.5).
  For our LyC leakers, the average $N_e$(O~{\sc ii})  is $\sim$ 415 cm$^{-3}$ and
average $N_e$(S~{\sc ii})  excluding J1154+2443 and J1442$-$0209 is
$\sim$ 400 cm$^{-3}$.  
  It follows from the above discussion that electron number densities in our 
LyC leakers are similar to the electron number densities in high-$z$ galaxies
and are considearbly higher than $N_e$(O~{\sc ii})~ =~30~cm$^{-3}$ and  
254~cm$^{-3}$ \citep{Harshan2020} typical for local SDSS (DR7) 
star-forming galaxies and SFGs at $z$~$\sim$~1.5, respectively.

\begin{figure}
  \includegraphics[angle=-90,width=0.99\linewidth]{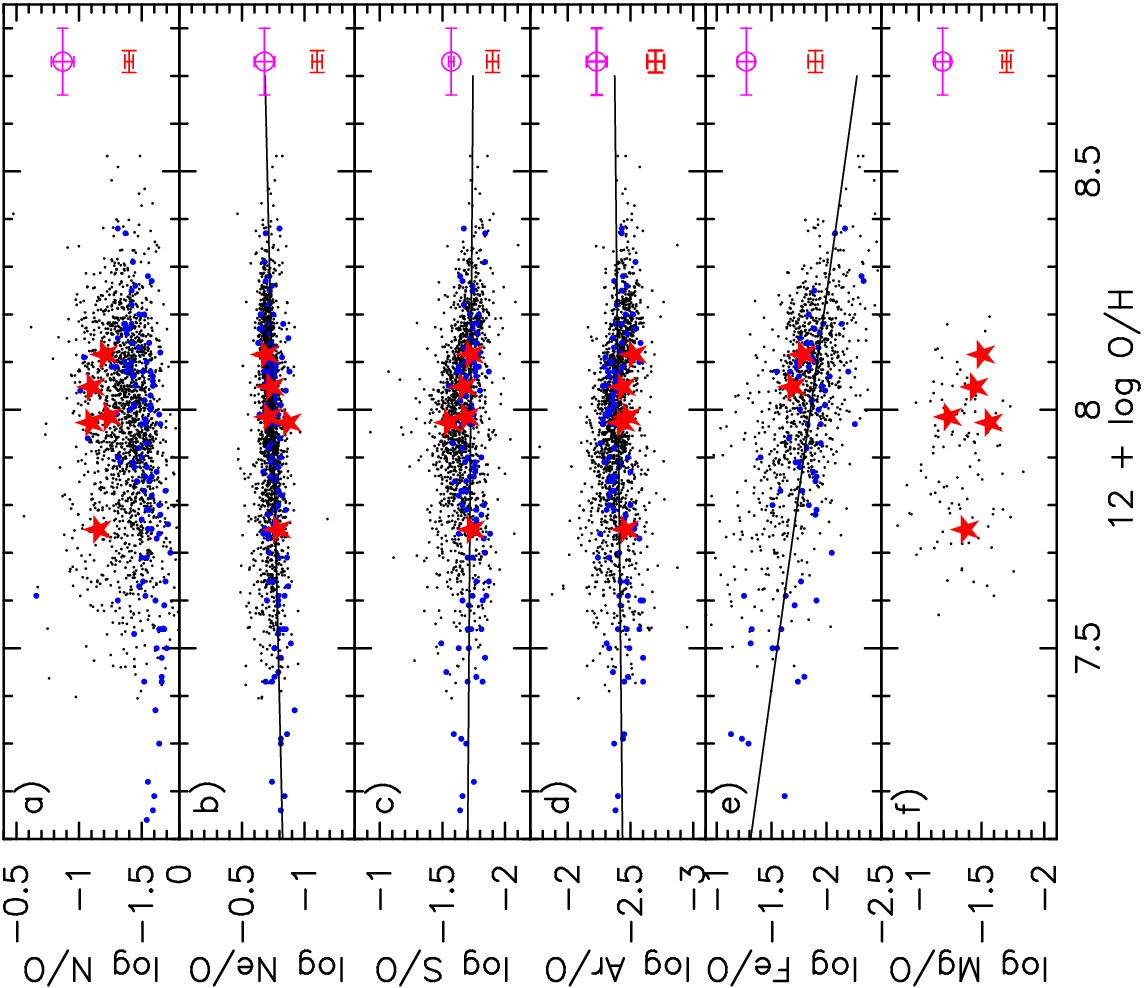}
  \caption{Dependences of different elemental abundance ratios X/O on the
oxygen abundance 12 + log O/H for our sample shown by red
stars. For comparison we show the HeBCD sample from \citet{IzTh2004}
and \citet{Izotov2004a}
used for the primordial helium abundance determination (blue circles),
and SFGs from the SDSS DR14 with the [O~{\sc iii}]4363\AA\ fluxes
measured with accuracy better than 4$\sigma$ (black dots). Regressions to 
the reference data are
presented by straight lines and solar values of \citet{Lodders2020} are drawn
by large magenta circles with error bars and averaged errors of LyC leakers
by error bars in red. 
}
  \label{fig4}
\end{figure}

\begin{figure}
\hbox{
\includegraphics[angle=-90,width=0.7\linewidth]{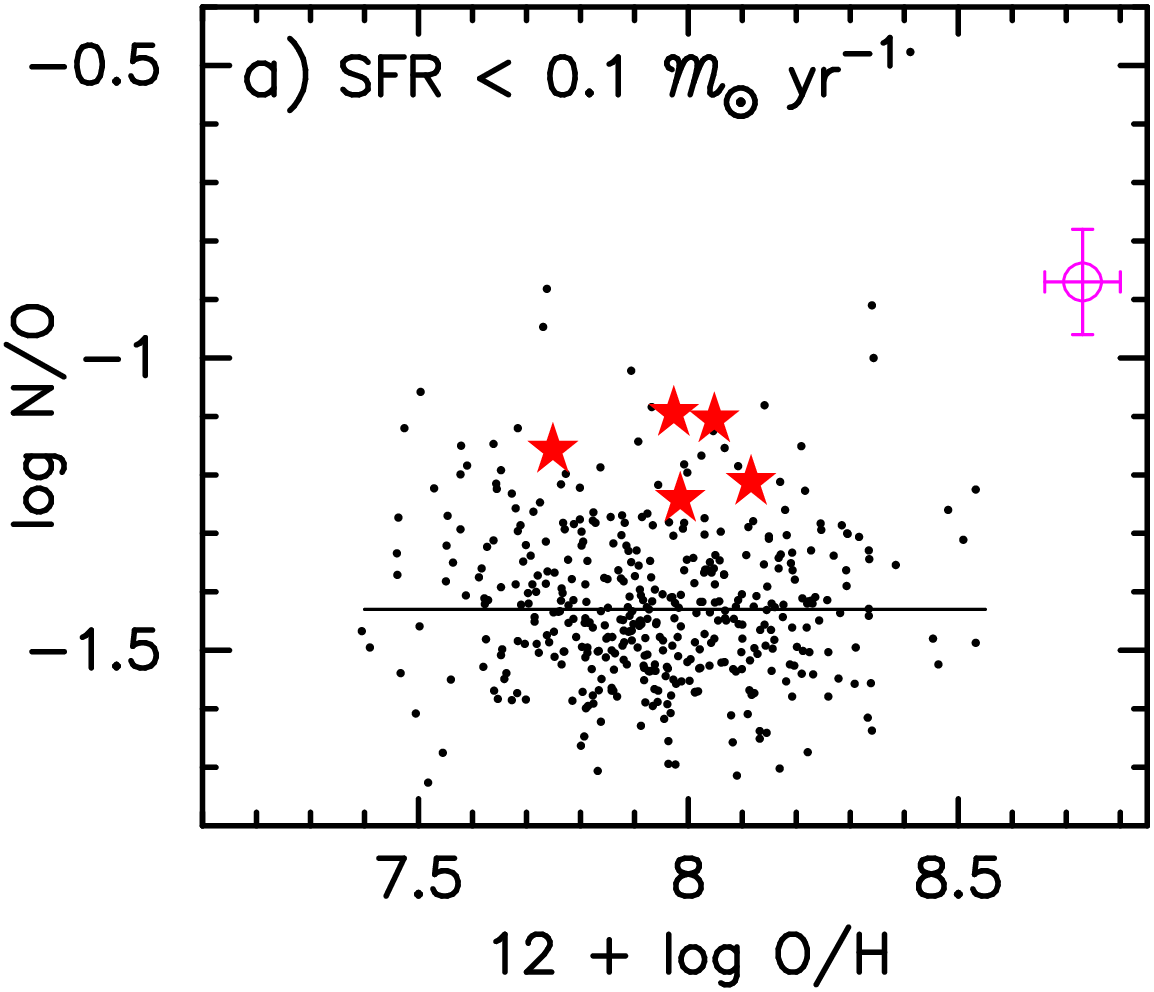}}
\vspace{0.3cm}
\hbox{
\includegraphics[angle=-90,width=0.7\linewidth]{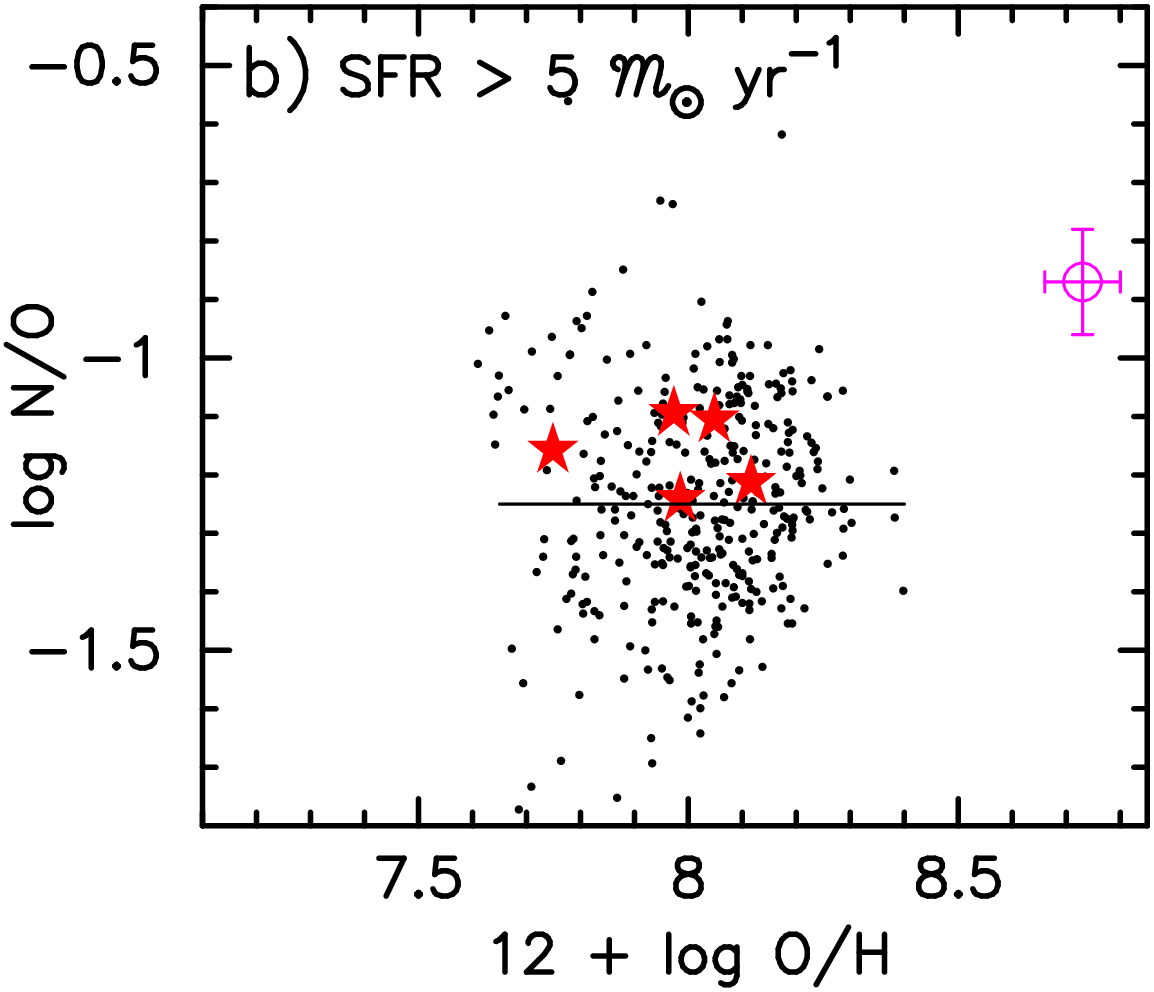}}
\vspace{0.3cm}
\hbox{
\includegraphics[angle=-90,width=0.7\linewidth]{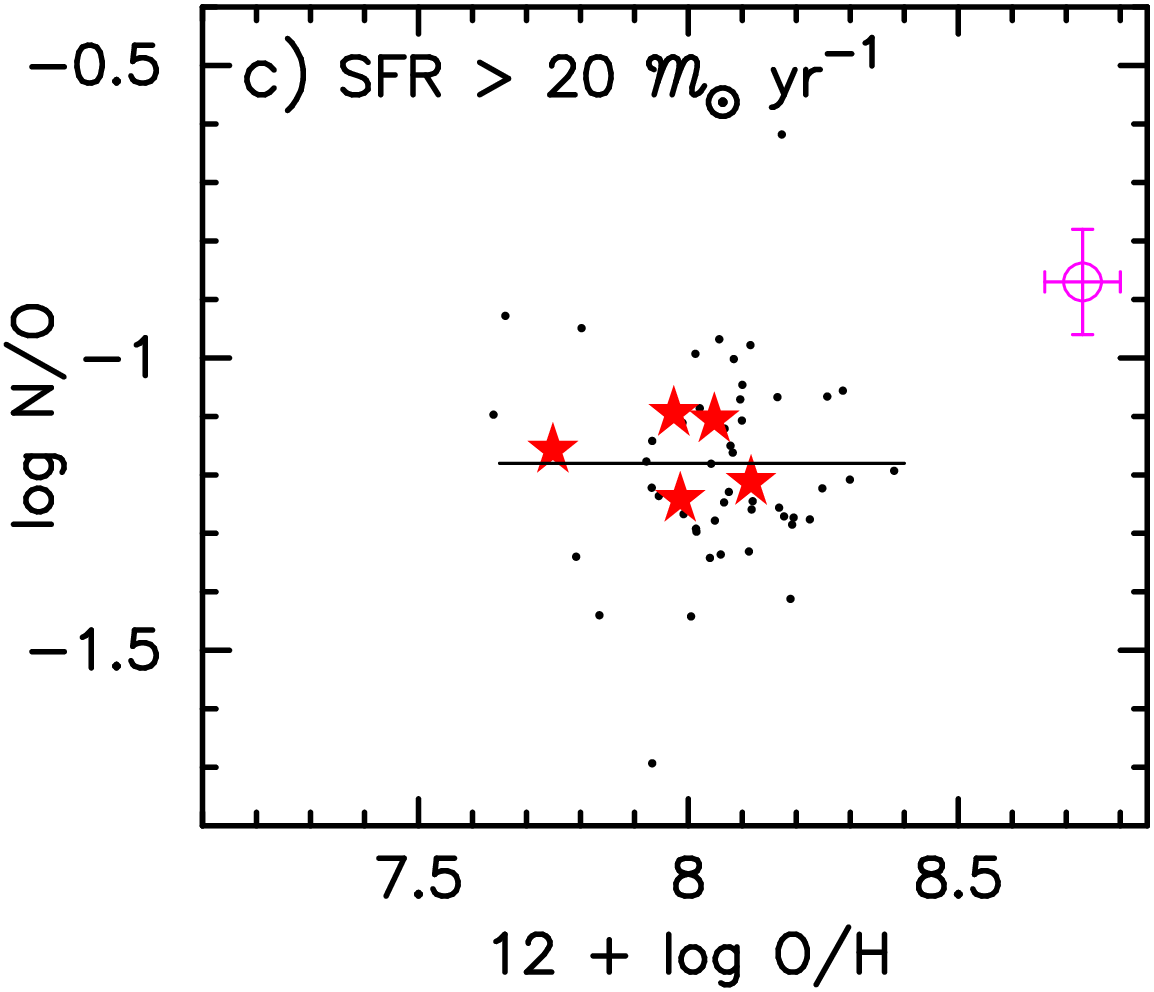}}
  \caption{Dependence of the N/O abundance ratio on 
12~+~log~O/H for three different ranges of SFR \citep[derived from the 
extinction-corrected H$\beta$ luminosities following ][]{Kennicutt1998} for 
our LyC leakers and SFGs from the SDSS DR14. 
Symbols and samples are the same as in  Fig.~\ref{fig4}. Mean values of 
log N/O for the respective ranges of SFR are denoted by horizontal lines.
}
  \label{fig4a}
\end{figure}

\begin{figure}
  \includegraphics[angle=-90,width=0.85\linewidth]{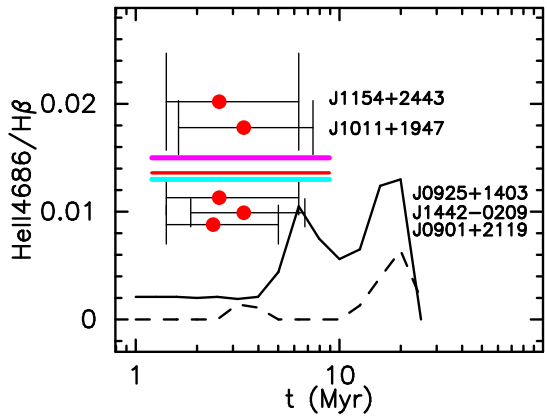}
\caption{Dependence of the He~{\sc ii}~4686\AA/H$\beta$ ratios on the 
starburst age.
CLOUDY v17.01 models in conjunction with the BPASS v2.1 stellar 
population models from \citet{Izotov2019}
for the instantaneous burst with the heavy element mass fraction of 10$^{-5}$
(10$^{-3}$) and nebular oxygen abundance 
12 + log O/H = 7.0 are shown by the solid (dashed) line. 
The observed XShooter He~{\sc ii}~4686/H$\beta$ flux ratios are shown by red
  filled circles with error bars.  
  Cyan and magenta lines indicate the position of local analogues of
high-redshift galaxies and low-redshift reference galaxies, respectively, both 
selected from SDSS by \citet{Bian2020} (see details in the text) together 
with the mean He~{\sc ii}/H$\beta$ ratio for our leakers (red line).
}
  \label{fig6}
\end{figure}

\begin{figure}
\hbox{
  \includegraphics[angle=-90,width=0.97\linewidth]{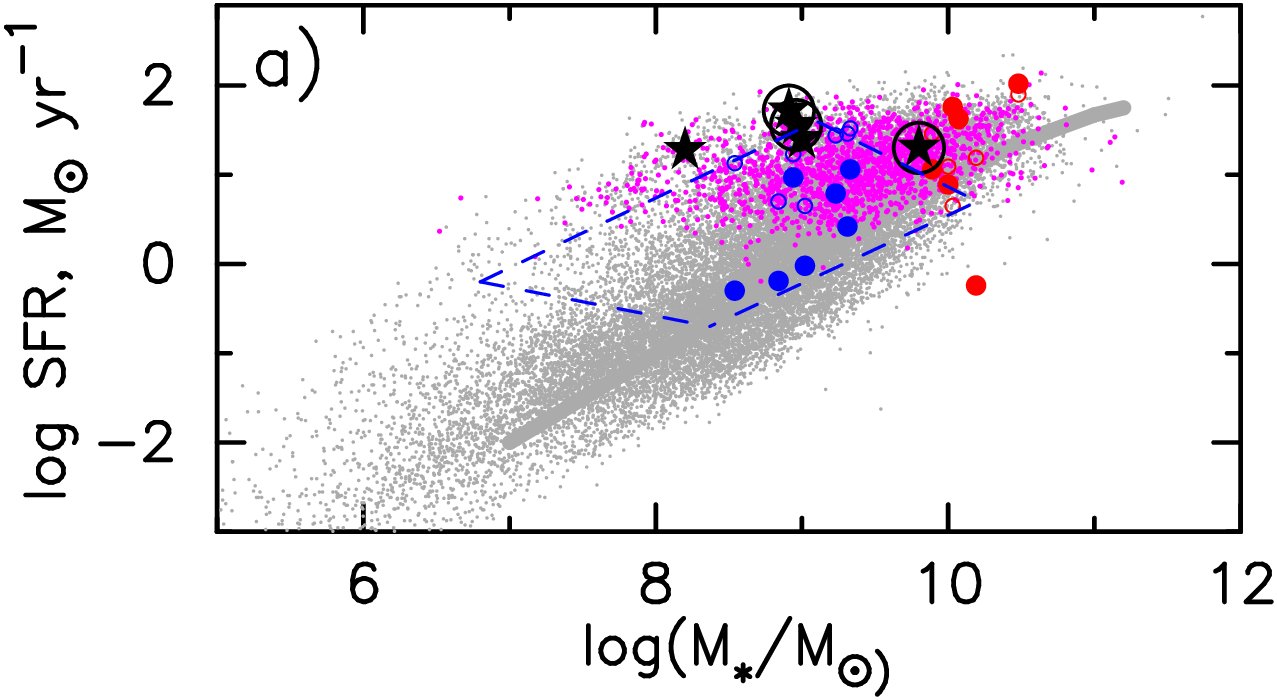}}
\vspace{0.3cm}
\hbox{
  \includegraphics[angle=-90,width=0.94\linewidth]{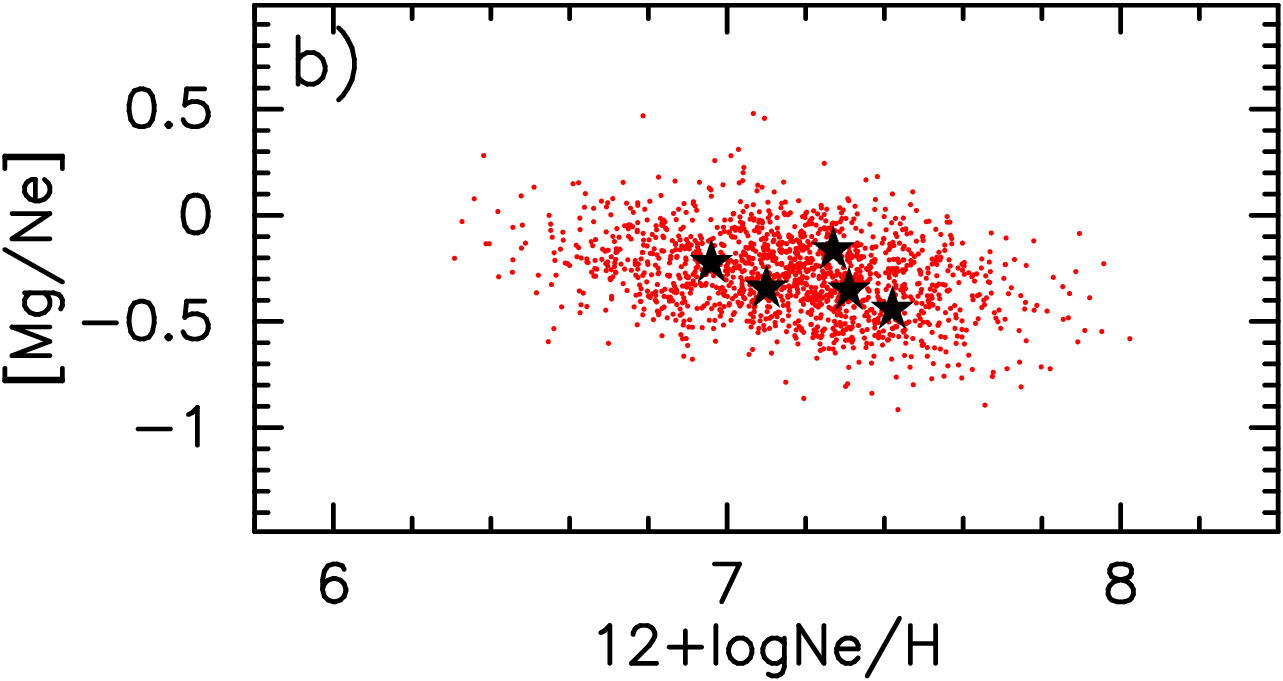}}
\vspace{0.3cm}
\hbox{
  \includegraphics[angle=-90,width=0.955\linewidth]{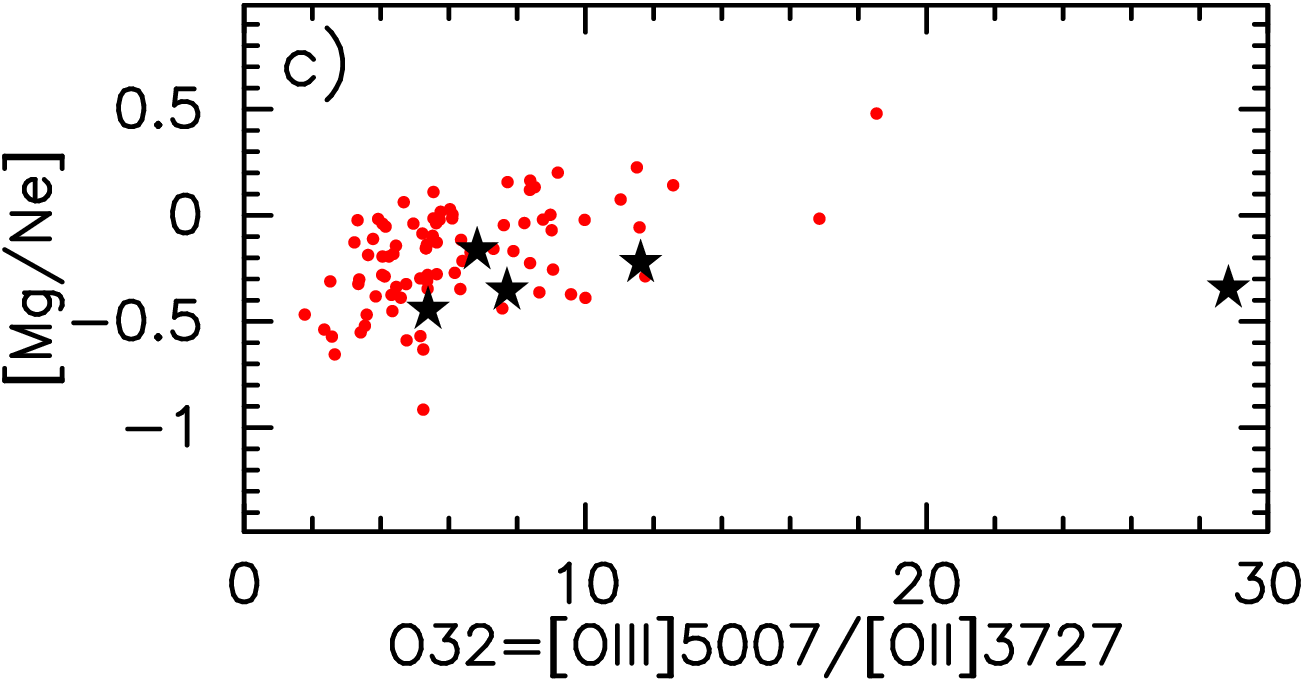}}
  \caption{{\textit a}) Dependence of SFR on stellar mass M$_\star$. 
Five our sample galaxies with
    Mg~{\sc ii} emission are shown by large black stars. Three galaxies with
    both Mg~{\sc ii} and Fe~{\sc ii}$^*$ emission in their spectra are 
    encircled by black circles.
      The entire Mg~{\sc ii} sample from \citet{Guseva2019}
    with high excitation H {\sc ii} regions with element abundances derived 
by the direct $T_{\rm e}$ method is shown by magenta dots.
      Additionally, SDSS DR14 SFGs  with
    EW(H$\beta$) $>$ 20\AA\ are shown by grey dots.
      The main sequence of SFGs of \citet{Finley2017}
    is indicated by thick grey line and their
    Mg~{\sc ii} and Fe~{\sc ii}$^*$ emitters 
    are shown by blue and red circles, where filled
    and open circles indicate galaxies with SFRs derived by different methods.
    The location of Mg~{\sc ii} emitters from \citet{Feltre2018} is indicated by
    dashed blue region.
         {\textit b}) Dependence of the magnesium-to-neon abundance ratio expressed
in [Mg/Ne] $\equiv$ log Mg/Ne -- log (Mg/Ne)$_\odot$ on the neon abundance 
12 + log Ne/H.
    Only galaxies from SDSS DR14 Mg~{\sc ii} sample with element abundances 
derived by the direct $T_{\rm e}$ method are shown (red dots).
         {\textit c}) Dependence of [Mg/Ne] on O32 = 
[O~{\sc iii}]~$\lambda$5007/[O~{\sc ii}]~$\lambda$3727. Symbols are the same as 
in {\textit b})
    but for Mg~{\sc ii} sample collected from DR14 only the galaxies with
    [O~{\sc iii}]~$\lambda$4363 intensity derived with accuracy 
    better than 
    4$\sigma$, with abundances derived by the direct $T_{\rm e}$ method
    and with EW(H$\beta$) $>$ 180\AA\ are shown (red dots).
}
  \label{fig7}
\end{figure}

   The dependences of the Ne/O, S/O, Ar/O, Fe/O and Mg/O ratios on oxygen abundance
for our LyC leakers are similar to those for other samples of
SFGs  \citep[e.g. HeBCD sample by ][]{IzTh2004,Izotov2004a} and 
SFGs from SDSS DR14
\citep*{IzStasMeynet2006a,IzGusevaT2011,Guseva2011,IzTGuseva2012}
with a similar spread of data (Fig.~\ref{fig4}).
   Note that in this Figure we only show SDSS DR14 
SFGs with precise data, where element abundances are derived by the direct
$T_{\rm e}$-method and the [O~{\sc iii}]~$\lambda$4363\AA\ fluxes are measured with an
accuracy better than 4$\sigma$.

\begin{table*}
  \caption{Ionic and Total Heavy Element Abundances \label{tab2}}
  \begin{tabular}{lrrrrr}  \hline \hline
  &\multicolumn{5}{c}{Galaxy}\\         
  {Property}&
\multicolumn{1}{c}{        J0901+2119} & \multicolumn{1}{c}{        J0925+1403} &
\multicolumn{1}{c}{        J1011+1947} & \multicolumn{1}{c}{        J1154+2443} &
\multicolumn{1}{c}{        J1442-0209}  \\ \hline
  $T_{\rm e}$(O {\sc iii}) (K) &
$13658\pm  232$ & $12426\pm  227$ &
$15142\pm  285$ & $16441\pm  505$ &
$14046\pm  278$  \\

  $T_{\rm e}$(O {\sc ii}) (K) &
$13144\pm  209$ & $12200\pm  210$ &
$14080\pm  247$ & $14720\pm  422$ &
$13410\pm  248$  \\

  $T_{\rm e}$(S {\sc iii}) (K) &
$12652\pm  193$ & $12135\pm  188$ &
$13677\pm  237$ & $15498\pm  419$ &
$12995\pm  231$  \\

$T_{\rm e}$(O {\sc iii}){\sc iraf}$^{\rm a}$ (K) &
$13777\pm  320$ & $12513\pm  306$ &
$15304\pm  396$ & $16646\pm  663$ &
$14173\pm  378$  \\

$T_{\rm e}$(O {\sc ii}){\sc iraf}$^{\rm a}$ (K) &
$13188\pm  662$ & $13594\pm  681$ &
 ...~~~~~       & ...~~~~~        &
$11894\pm  622$  \\
\\
$N_{\rm e}$(S {\sc ii}) (cm$^{-3}$) &
$  364\pm   92$ & $  225\pm   76$ &
$  608\pm  158$ & $  180\pm  150$ &
$   88\pm   66$  \\

$N_{\rm e}$(S {\sc ii}){\sc iraf}$^{\rm a}$ (cm$^{-3}$) &
$387\pm     50$ & $237\pm    42$  &
$652\pm    171$ & $173\pm   209$  &
$80\pm      38$   \\ 

$N_{\rm e}$(S {\sc ii})new$^{\rm b}$ (cm$^{-3}$) &
$298\pm     39$ & $194\pm    32$  &
$474\pm    114$ & $146\pm   168$  &  
$81\pm      35$   \\

$N_{\rm e}$(O {\sc ii}){\sc iraf}$^{\rm a}$ (cm$^{-3}$) &
$410\pm    14$  & $296\pm   13$   &
$547\pm    35$  & ${\bf 334}\pm   30$   &
${\bf 341}\pm    20$    \\

$N_{\rm e}$(O {\sc ii})new$^{\rm b}$ (cm$^{-3}$) &
$481\pm    17$  & $345\pm   15$   &
$646\pm    40$  & ${\bf 365}\pm   33$   &
${\bf 388}\pm    24$    \\
\\
  O$^+$/H$^+$ ($\times$10$^4$) &
$ 0.131\pm 0.007$ & $ 0.205\pm 0.013$ &
$ 0.034\pm 0.002$ & $ 0.048\pm 0.004$ &
$ 0.124\pm 0.008$ \\
  O$^{++}$/H$^+$ ($\times$10$^4$) &
$ 0.976\pm 0.051$ & $ 1.085\pm 0.063$ &
$ 0.890\pm 0.048$ & $ 0.502\pm 0.040$ &
$ 0.833\pm 0.048$ \\
  O$^{+++}$/H$^+$ ($\times$10$^6$) &
$ 0.896\pm 0.193$ & $ 1.497\pm 0.237$ &
$ 1.528\pm 0.239$ & $ 0.992\pm 0.241$ &
$ 1.017\pm 0.151$ \\
  O/H ($\times$10$^4$) &
$ 1.116\pm 0.051$ & $ 1.305\pm 0.064$ &
$ 0.939\pm 0.048$ & $ 0.561\pm 0.040$ &
$ 0.967\pm 0.049$ \\
  12 + log(O/H)  &
$ 8.048\pm 0.020$ & $ 8.116\pm 0.021$ &
$ 7.973\pm 0.022$ & $ 7.749\pm 0.031$ &
$ 7.985\pm 0.022$ \\
  \\
  N$^+$/H$^+$ ($\times$10$^6$) &
$ 0.115\pm 0.005$ & $ 0.135\pm 0.006$ &
$ 0.033\pm 0.002$ & $ 0.038\pm 0.002$ &
$ 0.078\pm 0.003$ \\
  ICF &
  7.617 &  5.932 &
  2.668 &  0.402 &
  7.103 \\
  log(N/O) &
$-1.106\pm 0.028$ & $-1.213\pm 0.029$ &
$-1.095\pm 0.033$ & $-1.158\pm 0.045$ &
  $-1.242\pm 0.030$ \\
  \\
  Ne$^{++}$/H$^+$ ($\times$10$^5$) &
$ 1.962\pm 0.114$ & $ 2.420\pm 0.159$ &
$ 1.273\pm 0.075$ & $ 0.872\pm 0.072$ &
$ 1.752\pm 0.112$ \\
  ICF &
  1.041 &  1.087 &
  0.985 &  1.038 &
  1.055 \\
  log(Ne/O) &
$-0.738\pm 0.034$ & $-0.696\pm 0.038$ &
$-0.874\pm 0.035$ & $-0.792\pm 0.049$ &
$-0.719\pm 0.038$ \\
  \\
  S$^{+}$/H$^+$ ($\times$10$^6$) &
$ 0.019\pm 0.001$ & $ 0.025\pm 0.001$ &
$ 0.005\pm 0.000$ & $ 0.008\pm 0.001$ &
$ 0.018\pm 0.001$ \\
  S$^{++}$/H$^+$ ($\times$10$^6$) &
$ 0.118\pm 0.009$ & $ 0.133\pm 0.011$ &
$ 0.067\pm 0.009$ & $ 0.052\pm 0.009$ &
$ 0.106\pm 0.009$ \\
  ICF &
  1.722 &  1.532 &
  3.611 &  1.694 &
  1.580 \\
  log(S/O) &
$ -1.675\pm 0.035$ & $ -1.731\pm 0.038$ &
$ -1.555\pm 0.056$ & $ -1.739\pm 0.070$ &
$ -1.693\pm 0.038$ \\
  \\
  Ar$^{++}$/H$^+$ ($\times$10$^7$) &
$ 0.033\pm 0.001$ & $ 0.034\pm 0.002$ &
$ 0.016\pm 0.001$ & $ 0.013\pm 0.002$ &
$ 0.026\pm 0.001$ \\
  Ar$^{+++}$/H$^+$ ($\times$10$^7$) &
$ 0.015\pm 0.003$ & $ 0.015\pm 0.003$ &
$ 0.022\pm 0.002$ & $ 0.023\pm 0.003$ &
$ 0.009\pm 0.002$ \\
  ICF &
  1.229 &  1.133 &
  2.114 &  1.428 &
  1.215 \\
  log(Ar/O) &
$-2.443\pm 0.050$ & $-2.530\pm 0.053$ &
$-2.436\pm 0.060$ & $-2.467\pm 0.126$ &
  $-2.482\pm 0.045$ \\
  \\
  Fe$^{++}$/H$^+$($\times$10$^6$)($\lambda$4658) &
$ 0.196\pm 0.035$ & $ 0.372\pm 0.047$ &
 ...~~~~~~~  &  ...~~~~~~~  &
 ...~~~~~~~  \\
  Fe$^{++}$/H$^+$($\times$10$^6$)($\lambda$4988) &
$ 0.222\pm 0.028$ & $ 0.179\pm 0.025$ &
 ...~~~~~~~  &  ...~~~~~~~  &
 ...~~~~~~~  \\
  ICF &
  10.779 &  8.158 &
   ...~~~~~~~  &  ...~~~~~~~  &
   ...~~~~~~~  \\
  log(Fe/O) ($\lambda$4658) &
$-1.724\pm 0.079$ & $-1.633\pm 0.059$ &
 ...~~~~~~~ &  ...~~~~~~~  &
 ...~~~~~~~  \\
  $[$O/Fe$]$ ($\lambda$4658) &
$ 0.304\pm 0.079$ & $ 0.213\pm 0.059$ &
 ...~~~~~~~ &  ...~~~~~~~ &
   ...~~~~~~~ \\
  log(Fe/O) ($\lambda$4988) &
$-1.669\pm 0.058$ & $-1.951\pm 0.065$ &
 ...~~~~~~~ &  ...~~~~~~~ &
 ...~~~~~~~ \\
  $[$O/Fe$]$ ($\lambda$4988) &
$ 0.249\pm 0.058$ & $ 0.531\pm 0.065$ &
 ...~~~~~~~ &  ...~~~~~~~ &
   ...~~~~~~~ \\
  \\
  Mg$^+$/H$^+$ ($\times$10$^6$) &
$ 0.271\pm 0.014$ & $ 0.357\pm 0.022$ &
$ 0.084\pm 0.006$ & $ 0.140\pm 0.011$ &
$ 0.414\pm 0.022$ \\
  ICF &
  14.478 &  11.353 &
  30.020 &  16.738 &
  13.365 \\
  log(Mg/O) &
$-1.454\pm 0.030$ & $-1.508\pm 0.034$ &
$-1.570\pm 0.033$ & $-1.379\pm 0.047$ &
$-1.243\pm 0.032$ \\  
\hline  \hline
  \end{tabular}

  \hbox{$^{\rm a}$Values obtained with {\sc iraf} routine \textit{temden}.}  
  \hbox{$^{\rm b}$Electron number density obtained from Eq.7 of \citet{Sanders2016}.} 
  \hbox{For J1154+2443 and J1442$-$0209 most reliable values of electron number densities are highlighted.}
\end{table*}

\subsubsection{Nitrogen abundance}

   The distribution of nitrogen to oxygen abundance ratio is a special case.
  It is seen in Fig. \ref{fig4}a that our LyC leakers occupy
the upper part of the N/O spread with the mean value log~N/O = --1.16 that is
$\sim$ 0.3 dex lower than the solar value log~(N/O)$_\odot$ = --0.87.
   We note that there are many other SFGs at oxygen abundances 
12~$+$~log~O/H~$\la$~8 in Fig.~\ref{fig4}a with an enhanced 
log N/O, while the lowest log N/O for these oxygen abundances in SDSS and HeBCD 
samples attain values as low as $\sim$ --1.7 to --1.6
\citep[see also ][]{Amorin2010,Amorin2012,Sanders2016,Vincenzo2016,Kojima2017}.   
This metallicity range is usually attributed to the primary N production. 
   At higher 12~$+$~log~O/H $>$ 8--8.5 both primary and
secondary mechanisms 
may be responsible for the
observed enhancement
\citep*{Charlot2001,Koppen2005,Molla2006,Pilyugin2012,AndrewsMartini2013}.   
   In earlier data releases of the SDSS
\citep[e.g. ][]{Izotov2004a,Amorin2010,Amorin2012}  only
very few SFGs with log N/O $>$ --1.4 were found at 12~$+$~log~O/H $\la$ 8.   
    Now we have collected more data in the SDSS DR14, resulting in much higher 
number of galaxies with large N/O at the low and extremely low metallicity end
\citep[see also ][]{Sanchez-Almeida2016}.
   This leads to a flatter dependence of N/O on the oxygen abundance.
Similar enhancements of N/O are found in other our local LyC leakers
\citep{I16a,I16b,I18a,Izotov2018b} as well as
in high-$z$ analogues of the galaxies during the epoch of reionization 
(EoR) at $z$ $\sim$ 6~-~10
\citep[e.g. LAEs and LBGs at $z$~$\sim$~2 of ][]{Sanders2016,Kojima2017}.

   Several mechanisms were proposed to explain this enhancement. 
   For example, \citet*{Amorin2010,Amorin2012} and \citet*{Loaiza2020} attribute
the N/O 
increase to a recent inflow of the relatively low-metallicity gas which 
substantially lowers the oxygen abundance.
   Another mechanism was proposed by \citet{IzStasMeynet2006a}. 
They showed that the local N/O enhancement in dense nitrogen-enriched ejecta 
from the winds of the evolved most massive stars (WR stars) can reach a 
factor of $\sim$~20 during the first 4 - 5 Myrs after onset of the star-formation
burst. Adopting the electron number density in the clumps 10 times higher
than in the ambient H~{\sc ii} region they estimated the apparent enhancement of
N/O in compact SFGs to a factor of $\sim$~2. 
    Our leakers belong to low metallicity, compact (exponential
disc scale length $\alpha$ $\sim$~1~-~1.5~kpc),
low-mass ($M_\star$~$\sim$~10$^8$~-~10$^9$M$_\odot$)  and relatively high density
galaxies with very young starbursts ($t$~$\sim$~3~-~4~Myr), thus the local N/O 
enhancement may be very pronounced.

 It is worth noting that no broad N~{\sc iii}~$\lambda$4640,
He~{\sc ii}~$\lambda$4686, C~{\sc iv}~$\lambda$5808 emission lines are seen in 
the XShooter and SDSS spectra, likely due to the young age of star
formation of the brightest burst and relatively low metallicity.
Besides that the signal-to-noise ratio in the continuum in the integral
spectra of our LyC leakers is too low to detect very weak WR features.

  Effects of differences in star formation rate (SFR) and in star formation history of
galaxies on the evolution of the relative abundance N/O during the long time 
period have been the subject of comprehensive studies
\citep[see e.g.][]{Molla2006}. 
  The position of our LyC leakers on the diagram N/O vs. 12~+~log~O/H 
is similar to the position of SF galaxies from SDSS DR14 with
high rate of star formation SFR $>$ 20 M$_\odot$ yr$^{-1}$ (Fig~\ref{fig4a}c).
   Note, that the ranges of SFR in the Figure were chosen arbitrarily to 
emphasize the effect of SFR on the N/O ratio for a sample of SDSS DR14 SFGs.
  On the other hand our leakers occupy the region above the relationship 
for SDSS SFGs with low SFRs $<$ 0.1 M$_\odot$ yr$^{-1}$ (Fig~\ref{fig4a}a).

\subsection{He {\sc ii} emission lines}\label{subsec:HeII}

   Nebular helium emission line He~{\sc ii} 4686\AA\ can be used as an 
indicator of hard ionising radiation with the energy above 4 Ryd.
   The strong He~{\sc ii} 
emission line often observed in galaxies with active star formation cannot 
always be explained by WR stars. These stars are detected only in a half of SFGs 
with the detected nebular He {\sc ii} emission 
\citep{GIT2000,TI2005,Kehrig2015,Kehrig2018}.

He~{\sc ii} 4686\AA\ emission lines with intensities above 2 per cent 
of H$\beta$ are observed in $\sim$ 10 per cent of SFGs by \citet{TI2005}. 
  Such intense emission cannot be explained by photoionisation models of 
H~{\sc ii} regions ionised by stellar radiation powered by ``normal'', known
stellar populations. This problem of nebular  He~{\sc ii} emission is well-known 
and has been widely discussed in the literature \cite[e.g.][]{Schaerer1996,GIT2000,Shirazi2012}.
Different possible explanations have been put forward, including ``uncommon'' 
stellar populations,
shocks and X-ray binaries \cite[see][]{Kehrig2015,Kehrig2018,Izotov2019,Schaerer2019,Plat2019,Szecsi2015,Gotberg2018,Bian2020} 
although a consensus has not yet been reached.

Nebular He~{\sc ii} emission, with intensity ratios $\sim 0.8-2$ per cent
of H$\beta$, is observed in the Xshooter spectra of our five targets. Although 
weaker than in some SFGs from the \citet{TI2005} sample,  these intensities 
are higher than predicted e.g.\ from BPASS models, even at very low
metallicities.
This is illustrated in Fig.~\ref{fig6}, where we show  stellar population models
taken from \citet{Izotov2019} 
   who considered BPASS v2.1 stellar models \citep{Eldridge2017}
with heavy element mass fraction 
of 10$^{-3}$ and 10$^{-5}$, and nebular oxygen abundance 12~+~log(O/H) = 7.0 
together with the CLOUDY v17.01 model calculations
for the instantaneous burst \citep{Ferland17} to obtain the dependence of the 
He~{\sc ii} $\lambda$4686/H$\beta$ emission line ratio with age.
   Stellar masses and ages of SFGs are derived from spectral energy
distribution (SED) fitting of the SDSS
spectra \citep*[see for more details, e.g. ][]{I16b}.
   Indeed, the He~{\sc ii} $\lambda$4686/H$\beta$ emission line ratios
in our LyC leakers are higher than those predicted by the models with heavy 
element mass fraction 10$^{-3}$ and even with 10$^{-5}$ (dashed and solid lines,
respectively, in Fig.~\ref{fig6}).

    We compare He~{\sc ii}~$\lambda$4686/H$\beta$ for our LyC leakers with that of the
local analogues of high-$z$ galaxies by \citet{Bian2020}.
   The mean value of log [N~{\sc ii}]~$\lambda$6584/H$\alpha$ for our LyC leakers
range from --1.9 to --1.4.
  We selected from \citet{Bian2020} stacked values of He~{\sc ii}~4686/H$\beta$
for local analogues of high-$z$ galaxies and for low-$z$ reference SF galaxies
only the data which are in the same range of [N~{\sc ii}]~$\lambda$6584/H$\alpha$.
   In Fig.~\ref{fig6} they are shown together with the mean
He~{\sc ii}~$\lambda$4686/H$\beta$ value for our leakers.
   All averaged values are nearly the same. Thus, the hard ionising radiation in
the high-$z$ analogues and our LyC leakers is similar to that in local 
reference SFGs.
   However, there is so far no indication for a possible link between the 
hardness of the ionising radiation and LyC leakage.

\begin{figure}
  \includegraphics[angle=-90,width=0.99\linewidth]{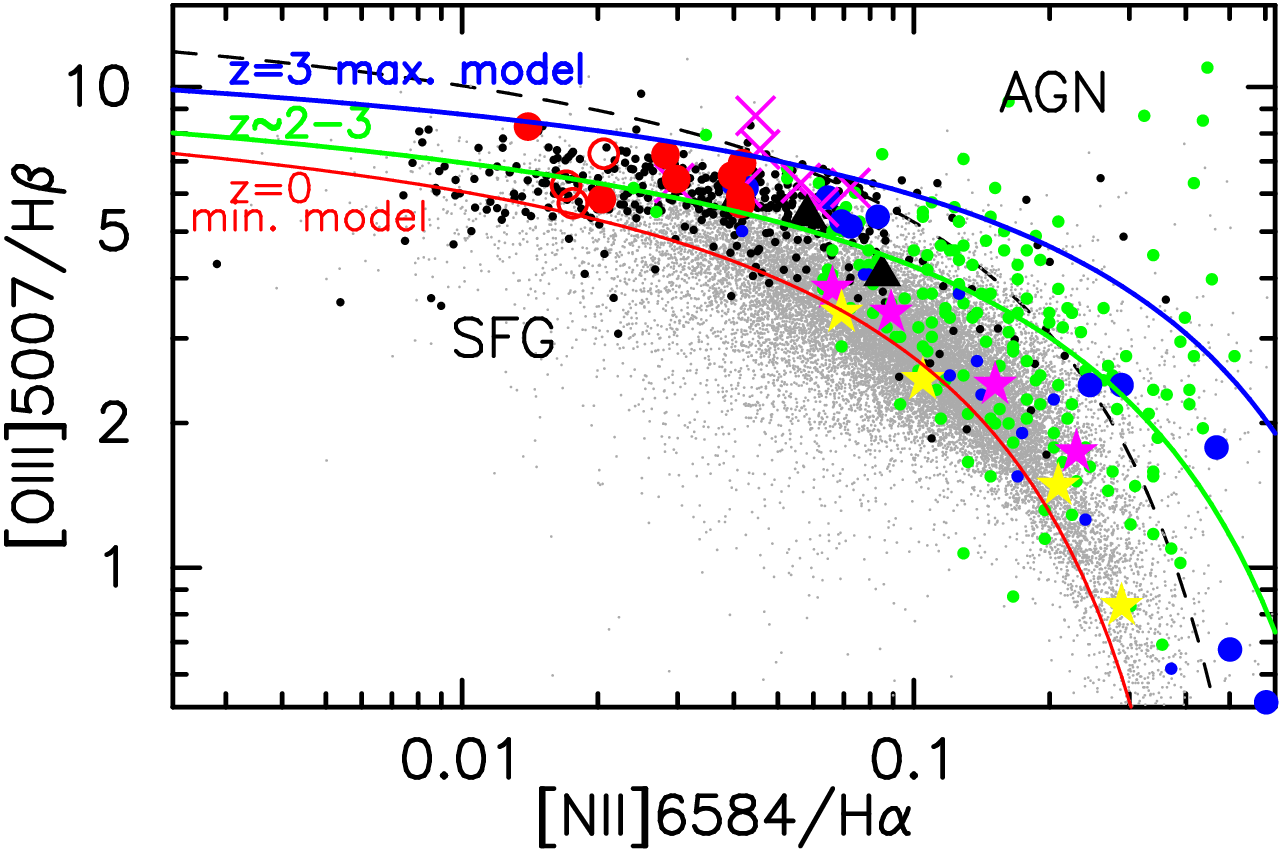}
  \caption{BPT diagnostic diagram \citep*{BPT81} for 11 confirmed LyC leakers
  discovered  by
    \citet{I16a,I16b,I18a,Izotov2018b} including the new
   XShooter measurements. The leakers with an accuracy better than 5$\sigma$ for the weak
   [N~{\sc ii}]~$\lambda$6584\AA\ emission lines 
   are shown by red filled circles, the remaining ones by red open circles.
   Nearly 30000 
   compact SFGs from SDSS DR14 are shown by grey dots from which
  the galaxies with large equivalent widths of H$\beta$
  (EW(H$\beta$) $>$ 180\AA) and accuracy of 
[N~{\sc ii}]~$\lambda$6584 flux measurements
better than 5$\sigma$ are presented by small black circles.
   Three GP galaxies by \citet{Amorin2012a} observed with GTC-OSIRIS are
presented by large black triangles. 
  Large blue circles are VLT/XShooter data for Lyman Break Analogues (LBAs)
from \citet{Loaiza2020} selected as having large offset in the BPT diagram.
    LBAs of \citet{Overzier2009} are denoted by small blue circles.
    We also show $z$ = 2 - 3 SFGs from KBSS-MOSFIRE sample
by \citet{Steidel2014} (green circles) and MOSDEF stacked galaxies from
\citet{Shapley2019} ($z$ $\sim$ 2 - 3, magenta stars and $z$ $\sim$ 1.5,
yellow stars).   
   UV-selected SFGs with the extreme emission line ratios at
$z$ $\sim$ 2 of \citet{Erb2016} are shown by magenta crosses.
   The separation line between SFGs and AGN by \citet{Kauffmann2003} is plotted 
as a dashed black curve.
   The minimum values (position of local galaxies at $z$ $\sim$ 0 along the
star forming branch) and  maximum values (SFGs at $z$ = 3) derived from
 theoretical models of \citet{Kewley2013a} are represented by red and blue
 lines, respectively.
    The best fit to the entire KBSS sample of SFGs is drawn by the green curve
    \citep[Eq.9 from ][]{Steidel2014}.
 }   
  \label{fig0}
\end{figure}

\begin{figure}
  \includegraphics[angle=-90,width=0.95\linewidth]{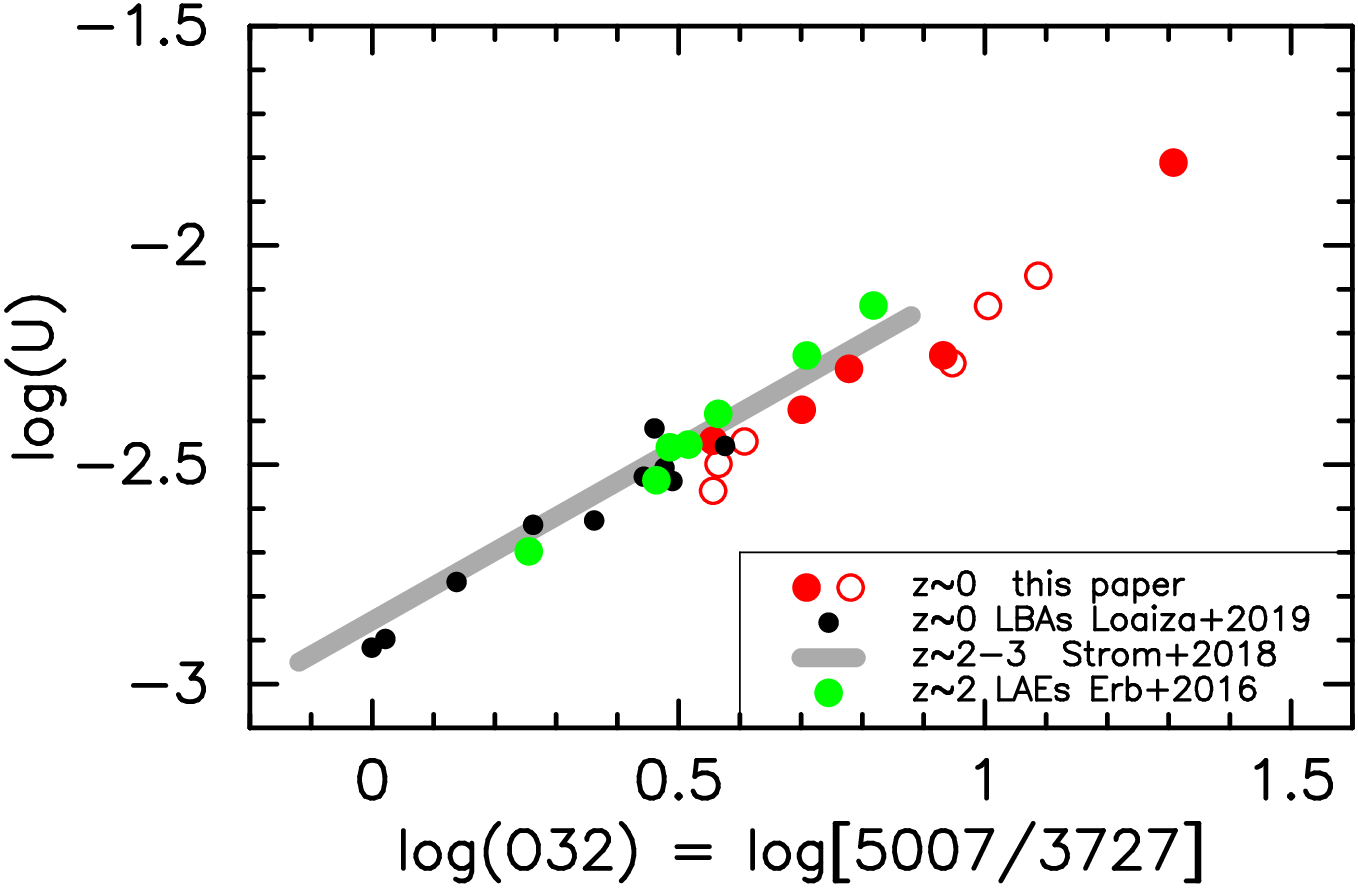}
  \caption{Relation between the ionisation parameter
    $U$ = $q$/$c$ estimated following  
\citet{KobulnickyKewley2004} and O32
for the confirmed LyC leakers discovered  by
\citet{I16a,I16b,I18a,Izotov2018b}  (red symbols, where filled circles denote 
objects for which metallicity is derived from XShooter data).
    Black circles are VLT/XShooter data for Lyman Break Analogues (LBAs)
from \citet{Loaiza2020}.
   LAEs from \citet{Erb2016} are denoted by green circles.
  The relation by \citet{Strom2018} derived for all
KBSS-MOSFIRE SFGs at $z$ $\sim$ 2~-~3 is presented by a straight line.
  }
  \label{fig00}
\end{figure}

\begin{figure}
\includegraphics[angle=-90,width=0.85\linewidth]{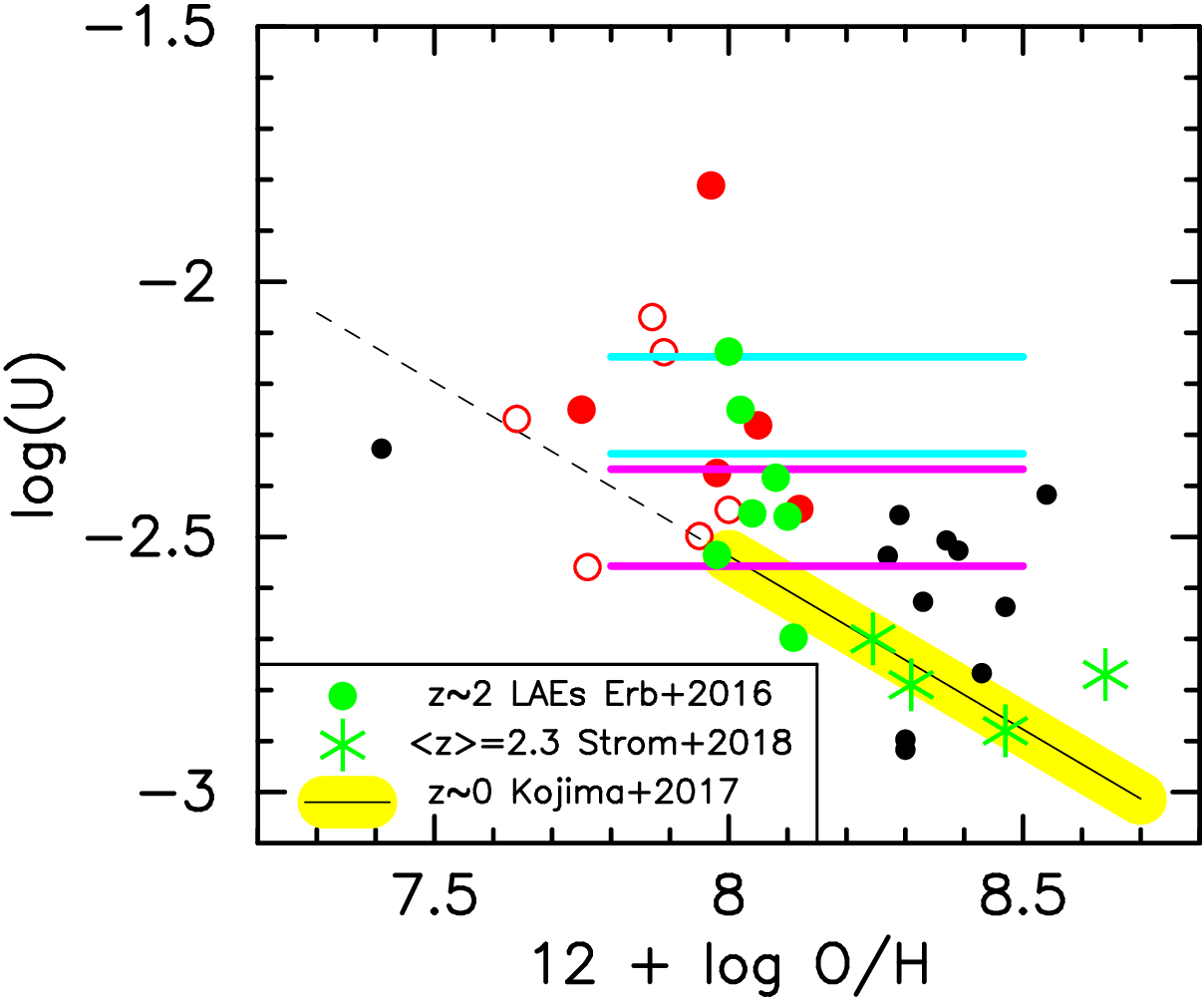}
  \caption{Dependence of the ionisation parameter
$U$ on 12~+~log~O/H for the same data as in Fig.~\ref{fig00}.
    Additionally, median values of $U$ in equal-number bins of 12~$+$~log~O/H for 148
$<$$z$$>$~=~2.3 KBSS galaxies from \citet{Strom2018} are shown by green asterisks.  
The relation by \citet{Kojima2017} (their Eq.\ 13) derived as the best fit to the
local $\sim$ 200,000 SFGs 
stacked in $M_\star$ and SFR by \citet{AndrewsMartini2013} is shown by
yellow and solid black lines. 
 The extrapolation of the best-fitting linear function of \citet{Kojima2017}
 to lower metallicities is denoted by a dashed line.
Cyan and magenta lines show the
position of local analogues of high-$z$ galaxies and low-redshift SDSS reference
galaxies, respectively, by \citet{Bian2020}. 
  }
  \label{fig01}
\end{figure}

\begin{figure}
\hbox{
\includegraphics[angle=-90,width=0.85\linewidth]{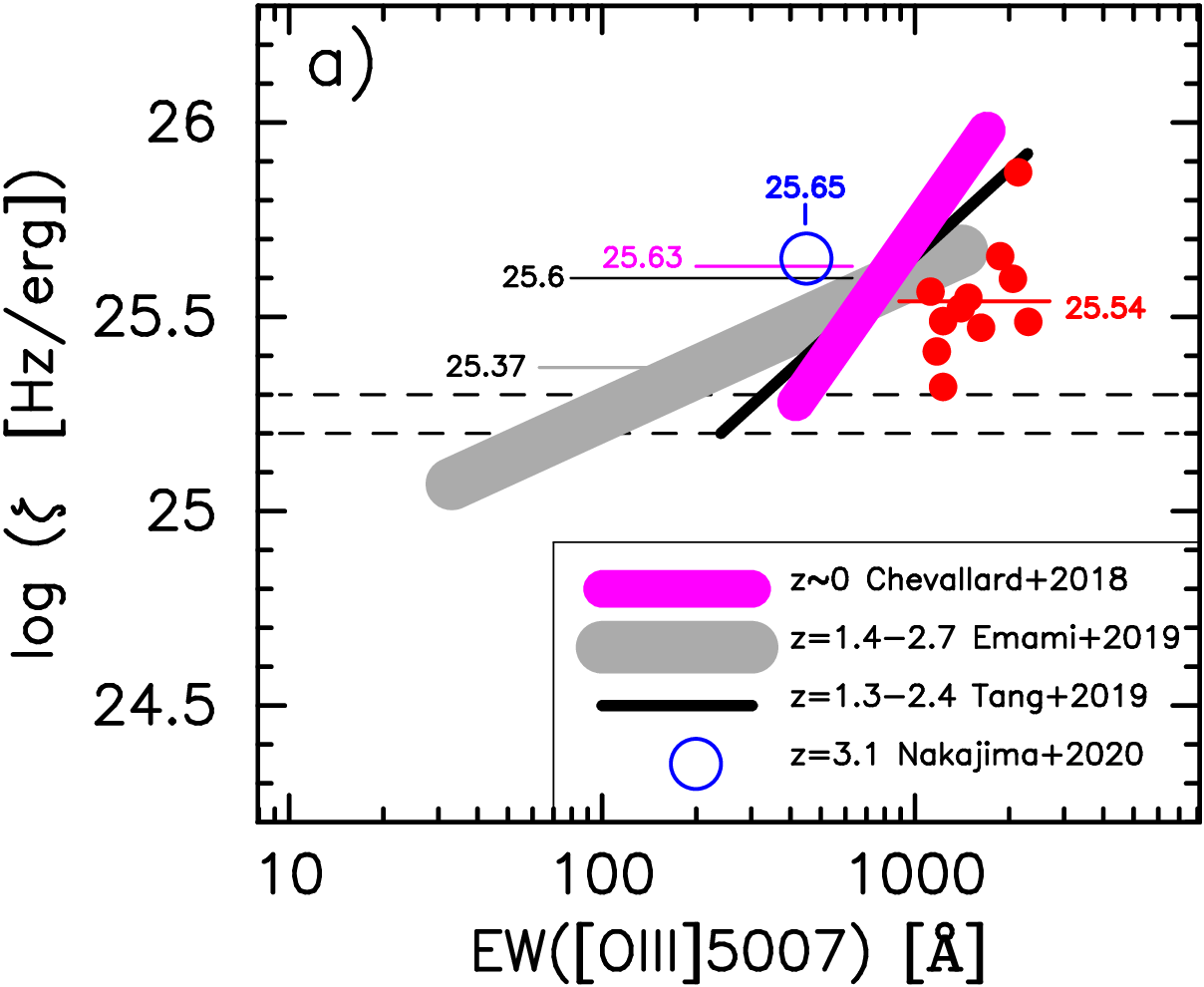}}
\vspace{0.3cm}
\hbox{
\includegraphics[angle=-90,width=0.85\linewidth]{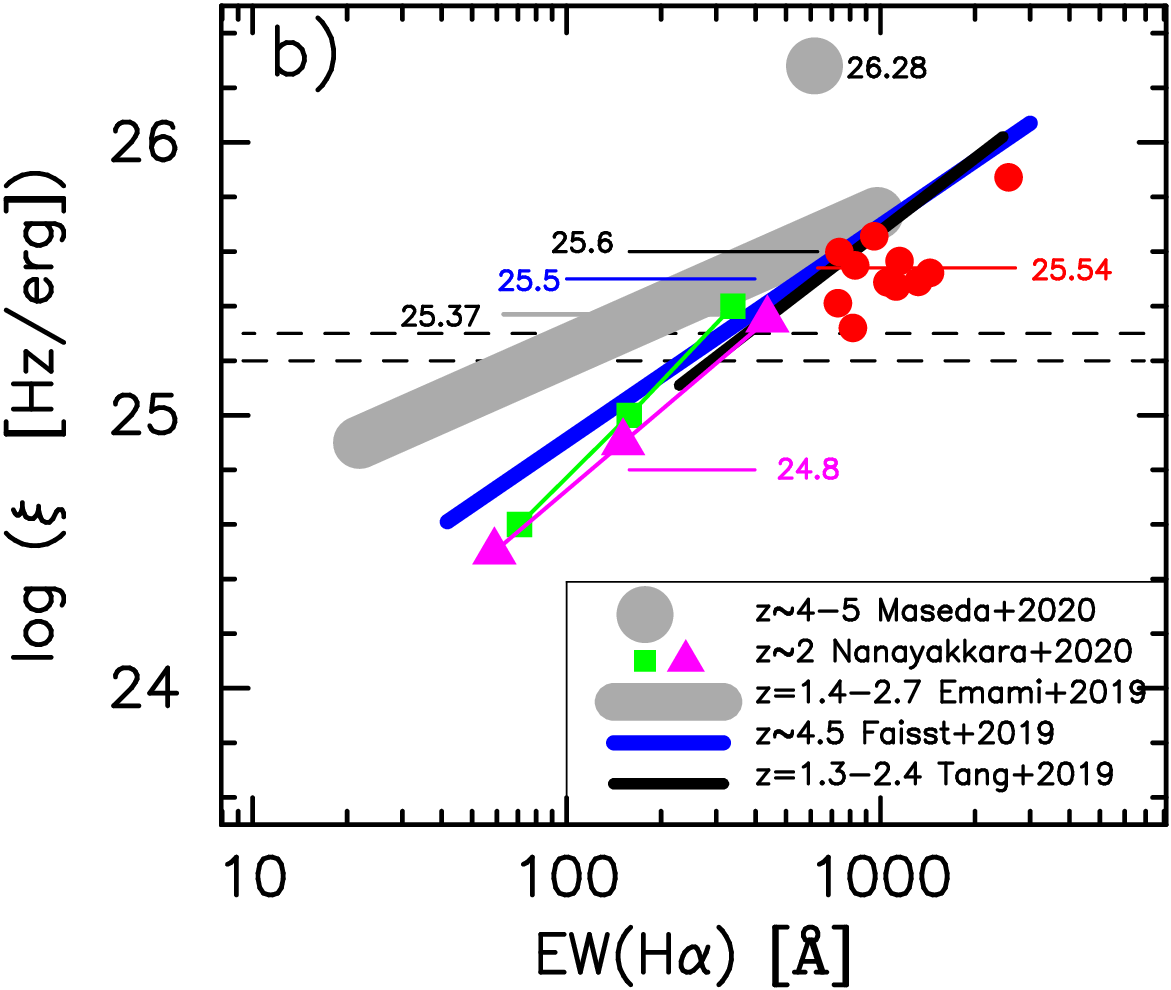}}
\caption{Relation between the ionising photon production efficiency $\xi$ 
  and EW([O~{\sc iii}]$\lambda$5007) (a) and EW(H$\alpha$) (b).
  Our LyC leakers are shown by red circles. The thick grey line in (a) is the best-fit to the sample of
 low-mass ($\sim$ 10$^{9.5}$ M$_\odot$) galaxies at $z$ = 1.4 -- 2.7 from
 \citet{Emami2019}. The position of extreme [O~{\sc iii}] emitters from
 \citet{Tang2019} is denoted by the light blue line.
 The LyC-LAE subsample (stacked LAEs with a clear LyC detection) of the LACES survey at
 $z$ $\sim$ 3.1 from \citet{Nakajima2020} is shown by the large open blue 
circle.
 The sample of local analogues of high-$z$ galaxies by \citet{Chevallard2018}
 is represented by the thick magenta line.
 (b). Data for high-$z$ galaxies of \citet{Emami2019}, \citet{Faisst2019}
  and \citet{Tang2019} are shown by thick grey,  blue and cyan
lines, respectively.
 Galaxies at $z$ $\sim$ 2 from \citet{Nanayakkara2020} are shown by
 green squares (entire sample) and magenta triangles (H$\beta$ detected 
galaxies).
Using stacked \textit{Spitzer}/IRAC photometry for $z$~$\sim$~4~-~5 galaxies
 \citet{Maseda2020} determined log~($\xi/{\rm [erg}^{-1} {\rm Hz]})$~=~26.28 (large grey open circle).
    The canonical values of log~($\xi/{\rm [erg}^{-1} {\rm Hz]})=25.2-25.3$ from \citet{Robertson2013} and
 \citet{Bouwens2016}, respectively, are shown by black dashed horizontal lines 
in (a) and (b).
   Average values of log~$\xi$ for different samples are denoted by 
corresponding positions of coloured horizontal lines and are labelled with the 
corresponding quantities.
}
  \label{fig02}
\end{figure}

\begin{figure}
  \includegraphics[angle=-90,width=0.95\linewidth]{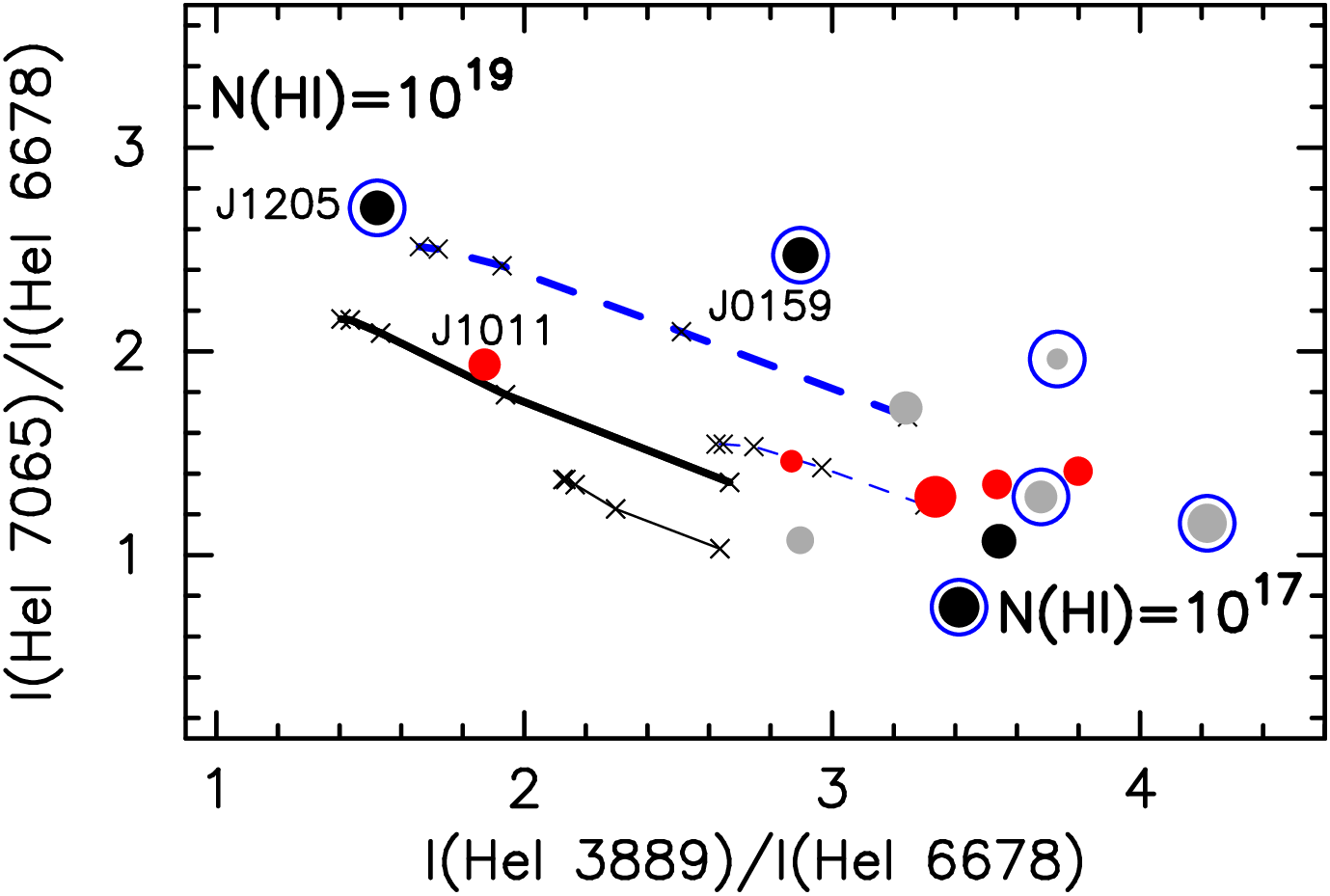}
  \caption{Diagnostic diagram for He~{\sc i}~$\lambda$3889/$\lambda$6678 vs.
  $\lambda$7065/$\lambda$6678 intensity ratios 
 proposed by \citet{ITG2017diag} for our
   sample shown by red circles. Confirmed LyC leakers from 
\citet{I16b,Izotov2018b} and Ly$\alpha$ emitting galaxies from 
\citet{Izotov2020} with He~{\sc i}
  fluxes measured from the SDSS spectra are shown by grey dots (the exception
  is J1333$+$6246 with a very noisy spectrum, which was discarded from the
  analysis).
   LBT spectrophotometric observations of the lowest-$z$ candidates to leakers 
with  extremely high O32 in the range of 23 - 43 \citep{ITG2017diag} are denoted
 by black symbols.
     Galaxies with oxygen abundances
12 + logO/H $<$ 7.75 are encircled with blue circles.
   All symbol sizes scale with
$f_{\rm esc}$(LyC) derived either through direct LyC observations
\citep{I16a,I16b,I18a,Izotov2018b} or using velocity separation between peaks,
$V_{\rm sep}$, in Ly$\alpha$ emission profiles \citep{Izotov2020}.
   CLOUDY models for a burst of SF with age $t$~=~2~Myr, fixed ionisation
parameter, filling factor log $f$ = --0.5 and oxygen abundance 12+logO/H = 7.3
(blue dashed) and 8.0 (black solid) and with two values of the electron number density
$N_{\rm e}$ = 1000 (thick) and 100 cm$^{-3}$ (thin) are shown by lines.
  Crosses on the lines mark increase of neutral hydrogen column densities from
$N$(H~{\sc i}) = 10$^{17}$ to 10$^{19}$ cm$^{-2}$ in 0.5 dex (from right to 
the left).
   }
  \label{fig5}
\end{figure}

\subsection{Mg~{\sc ii} and Fe~{\sc ii}$^*$ emission}\label{subsec:Mg}

   \citet{Henry2018} have shown that the Ly$\alpha$
escape fraction  in local compact SFGs tightly correlates with the
Mg~{\sc ii}~$\lambda$$\lambda$2796,2803\AA\ escape fraction. This implies that 
the Mg~{\sc ii}~$\lambda$$\lambda$2796,2803\AA\ emission lines can be considered
as a promising indicator of escaping Ly$\alpha$ and LyC emission.
The Mg~{\sc ii} emission lines are
observed in all five our galaxies. Additionally, non-resonant
Fe~{\sc ii}$^*$~$\lambda$2612 and $\lambda$2626 emission lines  are detected 
in three galaxies.
   We note that Fe~{\sc ii}$^*$  emission is observed in the galaxies with the strongest Mg~{\sc ii} emission (Table~\ref{tab1}). 
    Indeed, the average de-reddened  flux of Mg~{\sc ii}~$\lambda$2796 emission in 
our three galaxies with detected Fe~{\sc ii}$^*$ emission is $\sim$ 20 per cent of 
the H$\beta$ flux,
while the average flux of Mg~{\sc ii}~$\lambda$2796 emission in the two
galaxies without Fe~{\sc ii}$^*$ emission is $\sim$ 10 per cent of the H$\beta$.
A similar behaviour is found in $z$ = 1 - 2 galaxies by \citet{Erb2012}.
Mg~{\sc ii} emission with P Cygni profiles and Fe~{\sc ii}$^*$ emission
can be tracers of galactic outflows \citep[e.g. ][]{Finley2017}. This
emission may also originate in the H~{\sc ii} region
\citep{Guseva2013,Finley2017,Henry2018,Guseva2019}.
However, we do not detect P Cygni profiles in any emission lines including 
Mg~{\sc ii} lines.
   Instead, we detect signs of  high-velocity  winds created by massive stars and/or 
supernova remnants (SNRs) in all our galaxies as indicated by broad emission 
components underlying the bright hydrogen lines and brightest forbidden lines 
of ions of some heavy elements. 
A more detailed study of these features
will be discussed in a subsequent publication (Amor\'in et al., in prep.).

    Mg~{\sc ii} and Fe~{\sc ii}$^*$ emission in our confirmed
LyC leakers follows the same 
relations with global galaxy properties as other samples of
SFGs with Mg~{\sc ii} and Fe~{\sc ii}$^*$ detections
\citep{Finley2017,Feltre2018,Guseva2019}. 
   Our galaxies with very young starbursts (EW(H$\beta$) = 180 - 350\AA)
are located in the upper part of the SFR - $M_\star$ relation (Fig.~\ref{fig7}a).
  These galaxies (excluding J0901$+$2119) are among the Mg~{\sc ii}
galaxies that deviate most from the main star-formation sequence by
\citet{Finley2017}.
  For a given stellar mass, our galaxies tend to have high
SFRs (by more than one order of magnitude higher than that for main sequence
galaxies).

  The average stellar mass of three galaxies from the LyC leakers sample with 
the Fe~{\sc ii}$^*$ detection is 2.67 $\times$ 10$^9$ M$_\odot$ or $\sim$~5 times
higher than the average stellar mass of two galaxies without 
Fe~{\sc ii}$^*$ emission (5.8 $\times$ 10$^8$ M$_\odot$, see Table \ref{tab3}), 
similarly to \citet{Finley2017}, who found that
Fe~{\sc ii}$^*$ emission is preferentially seen in high-mass galaxies compared
to galaxies with only Mg~{\sc ii} emission.
  At the same time the average Lyman continuum escape fraction $f_{\rm esc}$(LyC) 
from the Fe~{\sc ii}$^*$ emitting high mass galaxies ($f_{\rm esc}$(LyC) = 5.9 
per cent) is $\sim$~5 times lower than that for non-Fe~{\sc ii}$^*$ emitting 
lower mass galaxies ($f_{\rm esc}$(LyC) = 28.7 per cent).
Here for averaging we have used the $f_{\rm esc}$(LyC) values obtained by 
\citet{I16a,I16b,I18a,Izotov2018b} from the \textit{HST} COS spectra. 

\citet{Guseva2019} considered  
the dependence of magnesium-to-neon abundance ratios
on metallicity for a sample of more than four
thousand Mg~{\sc ii} emitting low-metallicity SFGs extracted from the SDSS DR14 
and derived depletion of magnesium comparing this ratio with the solar
abundance ratio of this moderately 
refractory element to noble neon. They obtained 
[Mg/Ne] = log(Mg/Ne) -- log(Mg/Ne)$_\odot$ $\simeq$ --0.4 at solar 
metallicity. We wish to check their conclusion adding the more precise 
XShooter data.
   In Fig.\ref{fig7}b we plot 
the SDSS DR14 Mg~{\sc ii} sample
but including only galaxies with abundances derived with the direct $T_{\rm e}$ 
method and with [O~{\sc iii}]~$\lambda$4363 fluxes measured with accuracy better
than 
4$\sigma$.
The new data definitely follow the trend
depicted by DR14 Mg~{\sc ii} sample despite the smaller metallicity range
of the LyC leakers. Thus, 
the LyC leakers have the same Mg~{\sc ii} properties as a large Mg~{\sc ii}
sample with unknown LyC escape fractions.

    \citet{Nakajima2014}  proposed to use
O32 = [O~{\sc iii}]~$\lambda$5007/[O~{\sc ii}]~$\lambda$3727 as a parameter
indicating the fraction of escaping ionising radiation 
from density-bounded H~{\sc ii} regions. Since that time many attempts have 
been made to find correlations between these two parameters.  
    Both observational results
\citep{deBarros2016,Izotov2018b,Naidu2018,Izotov2020}
and model calculations \citep{Bassett2019,Katz2020} increasingly indicate that
O32 alone is an insufficient indicator of ionising radiation leakage.
  Therefore, it is necessary to look for other additional indirect indicators
of the LyC escape. Nevertheless, O32 is still potentially
useful, albeit it is an insufficient $f_{\rm esc}$(LyC) indicator.
    Therefore, it is of interest to study the behavior of this indirect
indicator in various dependences with other parameters.

  A comparison of the empirical dependences between O32 and [Mg/Ne] made 
by \citet{Guseva2019} with the data for our galaxy sample
is shown in Fig.~\ref{fig7}c.
   In the Figure only galaxies with the parameters closest to
those for our LyC leakers sample were selected from our DR14 sample, i.e. we used 
galaxies with EW(H$\beta$) $>$ 180\AA.
   In this case the mean value is [Mg/Ne]~$=-0.19$ for 86 DR14 Mg~{\sc ii} emitting galaxies.
If we restrict the selection from DR14 by the condition for the
EW(H$\beta$) $>$ 350\AA\ when dust grains likely would be destroyed by hard 
intense UV radiation,  the mean value of [Mg/Ne] for remaining nine DR14 
galaxies is near zero (--0.08) at its mean metallicity of 12 + logO/H = 7.8.
   As \citet{Izotov2011,Izotov2014b,Izotov2014c} showed, the presence of such
large EW(H$\beta$) leads to effective warming-up and destruction of
interstellar dust grains.
For LyC leakers the average [Mg/Ne] is equal to --0.31 at its mean
metallicity of 12 + logO/H = 8.0. 
  Given the similarity of metallicities and EW(H$\beta$) in the two
samples, the differences between mean [Mg/Ne] could be
explained by low neutral hydrogen column density
$N$(H~{\sc i}) in XShooter leakers or by possible different geometry
of the neutral gas distribution. 
   Note also, that the mean SFR of our LyC  leakers is 2 times higher than
that of the 9 galaxies from the DR14 sample. 

    Four of our galaxies with O32 $<$ 12 tentatively follow the trend derived 
from the DR14 sample while one galaxy with O32 $\sim$ 30 is located 
considerably lower compared to that expected from the trend.
  However, it must be emphasized that our XShooter sample is small to give
fairly definite conclusions.

\subsection{BPT diagram and related diagnostics}\label{subsec:BPT}

  Ongoing searches for the evolution of the properties of galaxies with
redshift led to the discovery of so-called an ``offset'' in the BPT diagram
\citep*{BPT81} between the locus of high-$z$ galaxies 
and that of typical local galaxies. 
  The interpretation of the BPT diagram (for example the O3N2 diagram) 
is also important in connection with applicability of the strong line methods developed for local
galaxies to determine metallicity and other physical parameters for distant
galaxies.
   Recent investigations of high-redshift galaxies based on large samples,
specifically on the
Keck Baryonic Structure Survey (KBSS) \citep{Steidel2014} and the MOSFIRE Deep
Evolution Field (MOSDEF) Survey \citep{Kriek2015}, have confirmed the offset
of high-$z$ galaxies compared to the typical local SFGs.

   Several explanations for such offsets have been put forward, including a higher ionisation parameter,
harder ionising spectra, higher electron number densities $N_{\rm e}$
\citep{Steidel2014,Hayashi2015,Strom2017,Strom2018,Shapley2019},
higher N/O abundance ratios at a given O/H in high-$z$ galaxies
\citep{AndrewsMartini2013,Masters2014,Kojima2017,Shapley2019,Loaiza2020},
 younger ages of the ionising population \citep[e.g.][]{Hayashi2015,Topping2019}, or combinations thereof.
Photoionisation models show that variations of one or several of these parameters
can in principle explain the observed offsets \citep{Kewley2013a,Steidel2014,Sanders2016}.
Although the most recent studies seem to favour harder ionising spectra or higher ionisation parameters,
no clear consensus on the dominant factor(s) has yet been reached \citep[see e.g.][]{Sanders2020,Bian2020}.

  We use the new XShooter observations of five confirmed LyC leakers with high
$f_{\rm esc}$(LyC) $\sim$ 3 - 46 per cent to construct the BPT diagram
with more accurate line intensity measurements. For the remaining galaxies from the
\citet{I16a,I16b,I18a,Izotov2018b} sample (6 galaxies, $f_{\rm esc}$(LyC) 
$\sim$ 2 - 73 per cent) the SDSS data were used.
  All LyC leakers with flux accuracy of
the weakest emission line [N~{\sc ii}]~$\lambda$6584\AA\ exceeding 5$\sigma$ are
emphasized by red filled circles, the rest galaxies are shown by 
open circles in Fig.~\ref{fig0}.
   The reference sample of $\sim$ 30000 compact SFGs from SDSS DR14 is shown by
grey dots.
   We note that  the 
H$\beta$ equivalent width in all 11 LyC leakers is high, 180 - 430\AA.
   Therefore, for a better comparison of our sample with the reference sample,
the SDSS SFGs with EW(H$\beta$) $>$ 180\AA\ are denoted in the Figure
by black dots.
  In principle, the hardness of ionising spectrum can be tested using the
[O~{\sc ii}]/[O~{\sc iii}] versus [S~{\sc iii}]/[S~{\sc ii}] relation, as
shown by \citet*{Perez2019}.
  However, these sulfur lines fall in the wavelength range of strong telluric
absorption lines distorting XShooter observed fluxes.

   For a comparison we also show local sample of Lyman Break Analogues (LBAs) by
\citet{Overzier2009}, LBAs observed with the VLT/XShooter by \citet{Loaiza2020},
high-redshift ($z$ = 2 - 3) SFGs from KBSS-MOSFIRE sample
by \citet{Steidel2014} and MOSDEF stacked galaxies from
\citet{Shapley2019}. 

   As shown in Fig.~\ref{fig0}, our local LyC leakers
($z$~$\sim$~0.3~-~0.4) are located in the upper part of a
SFG-branch of the BPT diagram, coinciding with the position of LBAs from
\citet{Overzier2009} and \citet{Loaiza2020} and that of high-$z$ SFGs from
\citet{Steidel2014} and \citet{Erb2016}.
   The distribution of the LyC leakers overlaps with the upper part of 
the distribution of compact
SDSS SFGs with EW(H$\beta$) $>$ 180\AA\ (black dots).
   All these galaxies are located in the region corresponding to high ionisation 
parameters and hard UV ionising radiation \citep{Kewley2013a,Steidel2014}, 
and are offset with respect to the position of the typical local SFGs.
   Our LyC leakers have also high ionisation parameters in the range of 
   log($U$) = --2.5 - --1.7,
relatively high N/O abundances,
   and high average ionising photon production 
efficiencies $\xi$~=~10$^{25.54}$~Hz~erg$^{-1}$ (see text below).
   Note, that the high-$z$ galaxies from \citet{Shapley2019} and the majority of 
 high-redshift galaxies from \citet{Steidel2014} are located in the
low-excitation part of the BPT diagram for SFGs.
     From the above discussion we conclude that the extreme galaxies both 
at low- and high-$z$ are located in the same upper part of the BPT diagram for SFGs,
 which indicates similar physical conditions.
From our data we cannot identify a single dominant mechanism which explains the observed offset
in the BPT diagram.

\subsubsection{Ionisation parameter}

   The ionisation state of the gas in galaxies is characterized by the 
ionisation parameter $q$
or by the dimensionless ionisation parameter $U$ = $q$/$c$ that is the ratio of 
the number density of ionising photons to the number density of hydrogen.  
The O32 ratio (here O32 is defined as [O~{\sc iii}]~$\lambda$5007/[O~{\sc ii}]~$\lambda$3727)
can serve as an observational indicator of the ionisation parameter.
  Since the oxygen abundances O/H of our LyC leakers were determined 
with high accuracy of 0.02 - 0.03 dex  (see Table~\ref{tab2}), we calculated 
$U$ following
\citet{KobulnickyKewley2004} (their Eq.~13), 
and using new strong line measurements and metallicities.
   In Fig.~\ref{fig00} the relation between $U$ and O32 for
these galaxies is shown by red filled circles, where the remaining LyC leakers
from \citet{I16a,I16b,I18a,Izotov2018b} are denoted by open red circles.
   They have a similar slope but consist of 
some galaxies with higher $U$ and O32 compared to the sequence 
outlined by local LBAs from \citet{Loaiza2020} (black circles) and $z$ $\sim$2
LAEs by \citet{Erb2016} (green circles).
     Higher $U$ for some of our LyC leakers is probably the result of 
our selection of candidates to the leakers based on high O32.
  Our sample has 
a lower metallicity, with an average value of 12 + log O/H = 7.9 for 11 LyC
leakers compared to 8.3 for the LBAs and slightly lower compared to 8.05
for the LAEs (Fig.~\ref{fig01}).

We have also compared the behaviour of the  ionisation parameter as a function of metallicity
between our local LyC leakers, $z$ $\sim$ 2 LAEs \citep{Erb2016}, local LBAs
\citep{Loaiza2020} and $\sim$ 200,000 local SFGs from SDSS DR7 of
\citet{Kojima2017} (Fig.~\ref{fig01}).
   We also show dependence of the ionisation parameter on nebular metallicity
for 148 high-$z$ KBSS galaxies by \citet{Strom2018} ($<$$z$$>$ = 2.3)
represented by median values in equal-number bins of the oxygen abundance
(green asterisks).
   Our data clearly show lower metallicities and higher ionisation parameters
compared to high-$z$ and other local samples of typical SFGs and slightly lower 
metallicities and higher ionisation parameters for some LyC leakers compared to 
$z$ $\sim$ 2 LAEs by \citet{Erb2016}.

\subsubsection{Ionising photon production efficiency $\xi$}

  The ionising photon production efficiency of galaxies 
$\xi$ is defined as the ratio of the number of ionising photons produced per 
unit time $N$(LyC) (production rate) to the intrinsic monochromatic UV 
luminosity $L_{\nu}$ per unit frequency, commonly measured at rest frame 
wavelength $\lambda$=1500\AA. The production rate is calculated as $N$(LyC) = 
2.1$\times$10$^{12}$~$L$(H$\beta$),
following \citet{Storey1995}. The uncertainty of the $\xi$ 
determinations in all our galaxies is less than 10 per cent.

$\xi$ is important to constrain the properties of galaxies during the
reionisation of the Universe.
   The efficiency of ionising photon production quantifies also 
   the relative amount of massive ionising stars with respect to the number of less massive non-ionising stars 
   present in a galaxy.
Together $\xi$, the Lyman continuum escape fraction, 
and the UV luminosity density allow one to calculate if the sources considered (galaxies)
are sufficient to reionise the Universe \citep[see e.g. ][]{Naidu2019}.

   In Fig.~\ref{fig02} we show the dependence of the ionising photon production
efficiency $\xi$ on EW([O~{\sc iii}]~$\lambda$5007) and on
EW(H$\alpha$) for our sample of 11 confirmed LyC leakers
\citep{I16a,I16b,I18a,Izotov2018b}. It is interesting to compare our data with
those for other local and high-$z$ samples.
   For example, 130 H$\alpha$ detections from ZFIRE survey using KECK/MOSFIRE gives a
median $\log(\xi/{\rm [erg}^{-1} {\rm Hz]})=24.8$ for galaxies at $z$ $\sim$ 2 within the mass range of
10$^9$ - 3$\times$10$^{11}$M$_\odot$ \citep{Nanayakkara2020}.
    \citet{Emami2019} derived $\log(\xi/{\rm [erg}^{-1} {\rm Hz]})=25.37$ 
for a mass-complete sample
(7.8 $<$ log ($M_\star$/M$_\odot$) $<$9.8) 
in the redshift range 1.4 - 2.7.
The sample of 221 $z$ $\sim$ 4.5 galaxies  at log $M_\star$/M$_\odot$ $>$ 9.7
\citep{Faisst2019}
spans a large range of $\xi$ with a median
$\log(\xi/{\rm [erg}^{-1} {\rm Hz]}) \sim 25.5$. This is $\sim$ 0.3 dex higher than the typically assumed
canonical value $\log(\xi/{\rm [erg}^{-1} {\rm Hz]})= 25.3$ \citep{Bouwens2016}.
    Extreme [O~{\sc iii}]  emitters by \citet{Tang2019} at $z$ $=$ 1.3 - 2.4
reach the largest values of  $\log(\xi/{\rm [erg}^{-1} {\rm Hz]}) \sim$ 25.6 for the galaxies with highest
EW(H$\alpha$) $>$ 1000 \AA.  
   The sample of ten extreme nearby SFGs by \citet{Chevallard2018}
   shows a similar slope and
range of $\xi$ and EWs of strong emission nebular lines as high-$z$ SF
galaxies.

The galaxies in our sample cover too small a range in EW(H$\alpha$) and
EW([O~{\sc iii}]~$\lambda$5007) to determine the slope of the relation,
but together with the galaxies from \citet{Chevallard2018} they follow 
a relatively steep
relation, close to the relation for high-$z$ galaxies.
   All our LyC leakers have $\xi$'s which are among the highest compared to the
high-$z$ samples \citep{Faisst2019,Tang2019,Emami2019,Nanayakkara2020} and
local ones \citep{Chevallard2018}. 
    This echoes \citet{Chevallard2018}'s conclusions that there is not strong
evolution 
of the relation between $\xi$ and EW(H$\beta$) or 
EW([O~{\sc iii}]~$\lambda$5007) over  $0 \la z \la 2$.
    The strong increase of the ionising photon production with increasing 
equivalent width of strong nebular emission lines means that extreme galaxies
such as compact SFGs with maximal EWs or their subset of LyC leakers
might be important
contributors to reionisation of the early Universe.

\subsection{He {\sc i} emission line diagnostic} \label{subsec:HeIdiagnostic}

   Recently \citet{ITG2017diag} proposed a new approach to estimate
the neutral hydrogen column density $N$(H {\sc i}) based on flux ratios of
He~{\sc i} emission lines $I$($\lambda$3889)/$I$($\lambda$6678) and
$I$($\lambda$7065)/$I$($\lambda$6678),
which helps to qualitatively conclude whether the galaxy can be a Ly$\alpha$ and 
LyC leaker.
     Low average H {\sc i} column densities, low column-density channels in dense environment, or a diversity of geometries
and neutral gas porosity (low H {\sc i} covering fractions), or a combination thereof
\citep{Kimm2019,Kakiichi2019,Gazagnes2020}
can be important for the escape of ionising photons.
   The proposed method has been 
tested on several galaxies \citep{Izotov2020}.
  The disadvantage of this method -- that it relies on emission lines of relatively low intensities -- 
can be circumvented by observing them with large
telescopes and sufficient S/N ratio, as the case for our VLT/XShooter
observations, where the S/N for the helium lines is $\sim$ 20 - 30.

   In Fig.~\ref{fig5} we show the relation between the He~{\sc i} emission line
ratios for our confirmed LyC leakers (XShooter data, red circles) and
for the LBT sample of lower-redshift ($z$~$<$~0.1) 
LyC leaker candidates of \citet{ITG2017diag} (black circles).
   Additionally, we present confirmed LyC leakers of
\citet{I16b,Izotov2018b} for which He~{\sc i} fluxes were measured from
the SDSS spectra (grey circles).
  For  the leaker candidates from \citet{ITG2017diag}, $f_{\rm esc}$(LyC) was 
estimated using velocity separation $V_{\rm sep}$ between the two 
peaks of Ly$\alpha$ emission line profiles \citep{Izotov2020};
galaxies where the Ly$\alpha$ line profile consists of the emission line
superimposed on broad absorption were excluded.

From Fig.~\ref{fig5}  we see that almost all the galaxies shown are located in 
the region of  low $N$(H~{\sc i}), according to the  CLOUDY models 
(solid and dashed lines). The lines shown are
 CLOUDY models for a burst of SF with age $t$ = 2 Myr, fixed ionisation
parameter, filling factor log~$f$~=~--0.5, oxygen abundances 12~+~logO/H~=~7.3
(blue dashed line) and 8.0 (black solid line), and with two values of the electron number
density, $N_{\rm e}$ = 1000 (thick lines) and 100 cm$^{-3}$ (thin lines).
    All model He~{\sc i} ratios in Fig.~\ref{fig5}
are calculated for static H~{\sc ii} regions. In the models with velocity
gradients 3889/6678 and 7065/6678 ratios would be respectively higher and
lower for a fixed $N$(H~{\sc i}), because of line Doppler broadening
leading to lower optical depth of the 3889 transition.

The galaxy  J1205$+$4551 is an exception.
   We note that this galaxy does not have a directly derived $f_{\rm esc}$(LyC),
but only its estimation by using the value of $V_{\rm sep}$
\citep[Eq. 2 in ][]{Izotov2018b}. 
   Moreover, this galaxy is one of the most deviating objects
in the relation between
$f_{\rm esc}$(Ly$\alpha$) and directly derived $f_{\rm esc}$(LyC)
\citep[see fig.8 in the paper of][]{Izotov2020},
and it is located on the relation f$_{\rm esc}$(Ly$\alpha$) = $f_{\rm esc}$(LyC).
  Its oxygen abundance 12 + log O/H = 7.46 is one of the lowest among all 
galaxies shown in Fig.~\ref{fig5}.
  Two other galaxies marked in the Figure, namely, J1011$+$1947 and
J0159$+$0751, 
imply $N$(H~{\sc i}) $\geq$ 5$\times$10$^{17}$ cm$^{-2}$.
   For the latter galaxy \citet{ITG2017diag} have derived the 
highest electron density $N_{\rm e}$(He~{\sc i}) = 2246 cm$^{-3}$ among the five 
compact SFGs with extremely high O32 observed with the LBT \citep{ITG2017diag}.
    J1011$+$1947 has the highest electron number density among our LyC leakers, 
both from sulfur and from oxygen determinations (Table \ref{tab2}).
    Such high electron densities should move the galaxies
upward following to the CLOUDY models.

   Therefore, according to the diagram in Fig.~\ref{fig5}, all our XShooter leakers 
as well as other confirmed LyC leakers
\citep{I16a,I16b,I18a,Izotov2018b} and the lowest-$z$ candidates 
\citep{ITG2017diag} excluding J1205$+$4551 indicate low H~{\sc i} column 
densities  implying leakage of 
LyC radiation from these galaxies, likely
due to their density-bounded H~{\sc ii} regions.
   For comparison, \citet{ITG2017diag} in their fig.~8 have shown the
relation between the $I$($\lambda$3889)/$I$($\lambda$6678) and
$I$($\lambda$7065)/$I$($\lambda$6678) ratios for non-LyCs \citep[unpublished
LBT data and data by][]{Yang2017}.
  All Ly$\alpha$ emitting galaxies from \citet{Yang2017} and the vast
majority of
BCDs with O32 $\sim$ 5 - 20, observed with the LBT, are located in the region of
high $I$($\lambda$3889)/$I$($\lambda$6678)
and low $I$($\lambda$7065)/$I$($\lambda$6678) ratios corresponding to the case 
of low column density regions.
   However, there may be exceptions when ionisation-bounded galaxies with
very high ionisation parameter
have also high 
average $N$(H~{\sc i}), as it likely is in the case of J1205+4551, but
low-density channels are present through which the ionising
radiation can escape \citep{Kimm2019,Kakiichi2019}. 
   This suggests that the diagnostic method proposed
by \citet{ITG2017diag} can effectively indicate LyC leakage.

\section{Summary}\label{subsec:summary}  

We have obtained new VLT/XShooter spectra of five confirmed LyC leakers 
at $z \sim 0.3-0.4$  discovered recently with the {\sl HST}.
Using the spectra we study the physical properties 
of these compact star-forming galaxies, which have many similar parameters with 
extreme high-$z$ galaxies responsible for reionisation of the early Universe.
  Our main findings are the following:

  1. Our XShooter LyC leakers have the same distributions of Ne/O, S/O, Ar/O, Fe/O and Mg/O
abundance ratios with oxygen abundance as other samples of SF galaxies.
   An exception is the N/O abundance ratio, which is enhanced compared to
the bulk of nearby SFGs in the oxygen
abundance range 12 + log O/H $\simeq$ 7.7 to 8.1 with the mean
value of log N/O = --1.16. This value is similar 
to that found in other local and high-$z$ analogues of galaxies of the epoch of
reionisation.

2.  We find mean electron densities $N_e \sim 400$ cm$^{-3}$, 
which are significantly higher (by a factor of 10 or more) than those typical for local SDSS
star-forming galaxies, and are comparable to those measured in star-forming galaxies at $z \sim 2-3$.

  3. We detect  the nebular He~{\sc ii}~$\lambda$4686\AA\ emission line in all our galaxies with 
intensities of 1 - 2 per cent of the H$\beta$ emission line, indicating a relatively hard ionising spectrum
in these galaxies, comparable to other galaxies at the same metallicity.

  4. Strong resonant Mg~{\sc ii}~$\lambda$$\lambda$2796,2803\AA\
emission lines with $I(\lambda$2796,2803)/$I$(H$\beta$) $\simeq$ 10 - 38 per 
cent and without P Cygni features are observed in all five our galaxies.
   Additionally non-resonant 
Fe~{\sc ii}$^*$~$\lambda$2612 and $\lambda$2626 lines are detected 
in three of the five galaxies with stronger Mg~{\sc ii} emission, higher
$M_\star$ and with $f_{\rm esc}$(LyC)~$\sim$~5 times lower than that for 
non-emitting Fe~{\sc ii}$^*$ galaxies.
  Our LyC leakers show the same trend of [Mg/Ne] with metallicity
as an entire DR14 Mg~{\sc ii} sample exhibiting the same magnesium depletion
in dust.

   5. The location of the 11 compact $z \sim 0.3-0.4$ LyC leakers (from which our sample is drawn) 
   with EW(H$\beta$) = 180 - 430\AA\ in the BPT
diagram coincides with the upper boundary of compact SDSS SFGs with
EW(H$\beta$) $>$ 180\AA, positions of local LBAs
\citep{Overzier2009,Loaiza2020} and $z$ $\sim$ 2 - 3 SFGs
\citep{Steidel2014,Erb2016}.
   Our LyC leakers have high ionisation parameters 
log($U$) = --2.5 to --1.7 and high average ionising photon production 
efficiencies $\xi$ = 10$^{25.54}$ erg$^{-1}$ Hz .
   Therefore, we conclude that the extreme galaxies at both low- and high-$z$ reside
in the same part of the BPT diagram and have properties very similar
to most extreme $z$ $\sim$ 2 - 3 galaxies.

6.  Using new measurements of faint He~{\sc i} lines, we confirm the effectiveness of the He~{\sc i} emission line diagnostic
proposed by \citet{ITG2017diag} to identify LyC leaking galaxies.
   All the LyC leakers from \cite{I16a,I16b,I18a,Izotov2018b}
and the low-$z$ LyC leaker candidates of
\cite{ITG2017diag} except one (J1205$+$4551) are located in the region of low
neutral hydrogen column density $N$(H~{\sc i}),
which indicates leakage of LyC radiation from these galaxies.

\section*{Acknowledgements}

We are grateful to anonymous referees for useful comments on the manuscript.
N. G. G. and Y. I. I. acknowledge support from the National Academy of Sciences
of Ukraine by its priority project ``Fundamental
properties of the matter in the relativistic collisions of nuclei and in the
early Universe'' (No. 0120U100935).
J. V. M., E. P. M., J. I. P. and C. K. acknowledge financial support from the 
State Agency for Research of the Spanish MCIU through the "Center of Excellence
Severo Ochoa" award to the Instituto de Astrof\'{\i}sica de
Andaluc\'{\i}a (SEV-2017-0709) and project A2016-79724-C4-4-P.
R. A. acknowledges support from FONDECYT Regular 1202007.~~
Funding for the Sloan Digital Sky Survey IV has been provided by
the Alfred P. Sloan Foundation, the U.S. Department of Energy Office of
Science, and the Participating Institutions. SDSS-IV acknowledges
support and resources from the Center for High-Performance Computing at
the University of Utah. The SDSS web site is www.sdss.org.
SDSS-IV is managed by the Astrophysical Research Consortium for the 
Participating Institutions of the SDSS Collaboration. 
This research has made use of the NASA/IPAC Extragalactic Database (NED), which 
is operated by the Jet Propulsion Laboratory, California Institute of 
Technology, under contract with the National Aeronautics and Space 
Administration.

\section*{Data availability}.

The data underlying this article are available in the article and in its online
supplementary material. This paper has been typeset from a TEX/LATEX file
prepared by the author.

\appendix

\section{XShooter spectra}

\begin{figure*}
\includegraphics[angle=-90,width=0.85\linewidth]{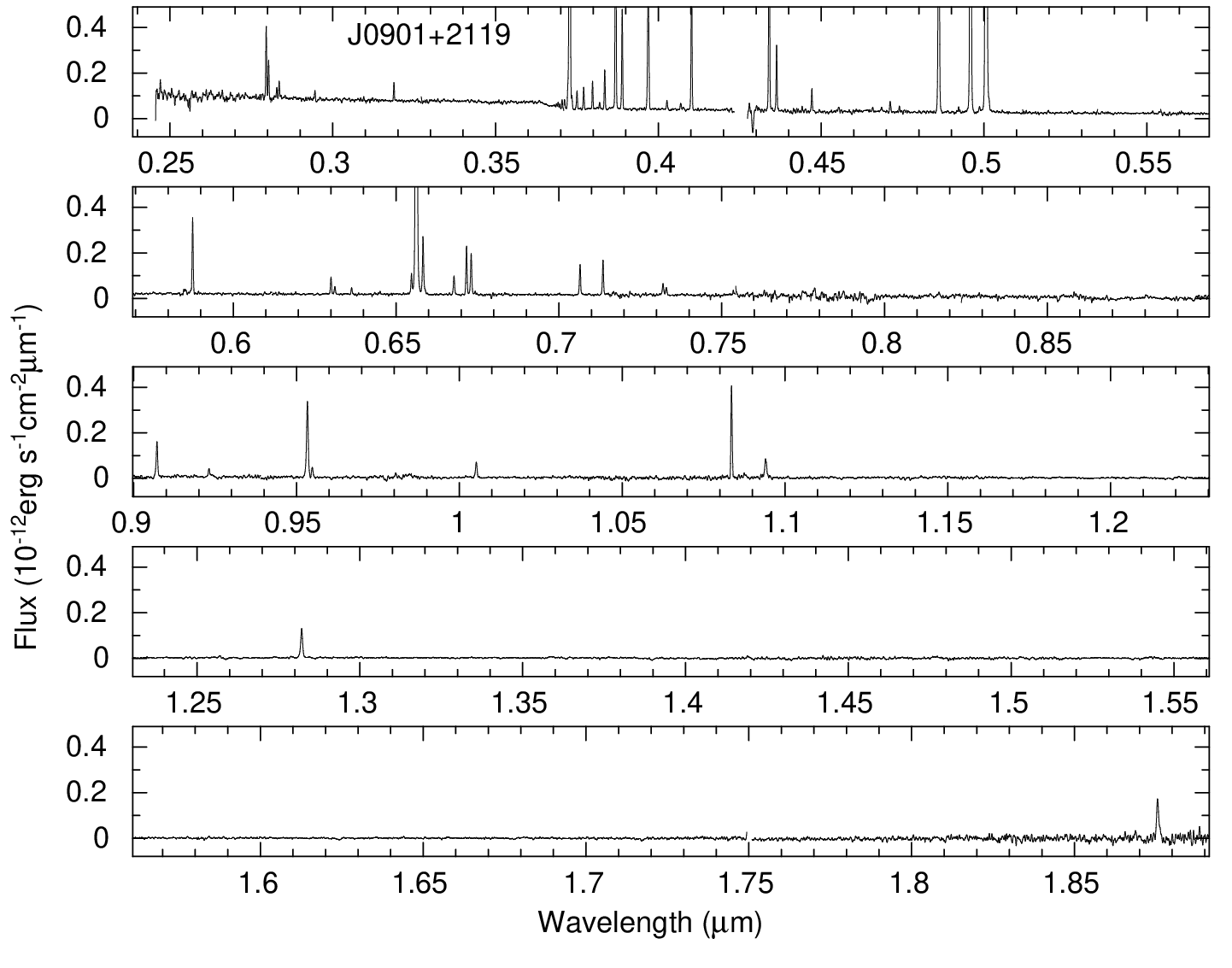}
\includegraphics[angle=-90,width=0.85\linewidth]{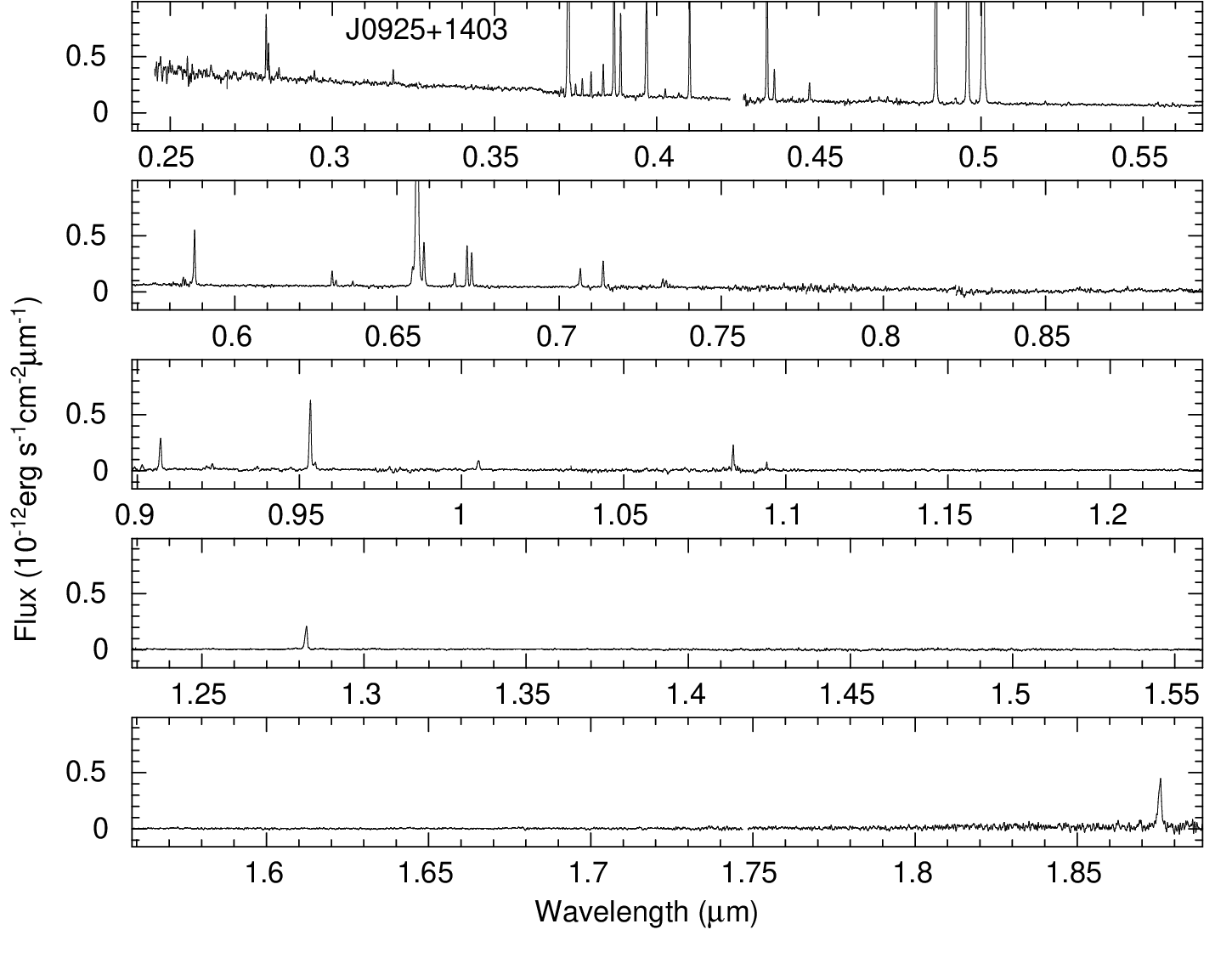}
\caption{Rest-frame XShooter spectra.}
\end{figure*}

\setcounter{figure}{0}

\begin{figure*}
\includegraphics[angle=-90,width=0.85\linewidth]{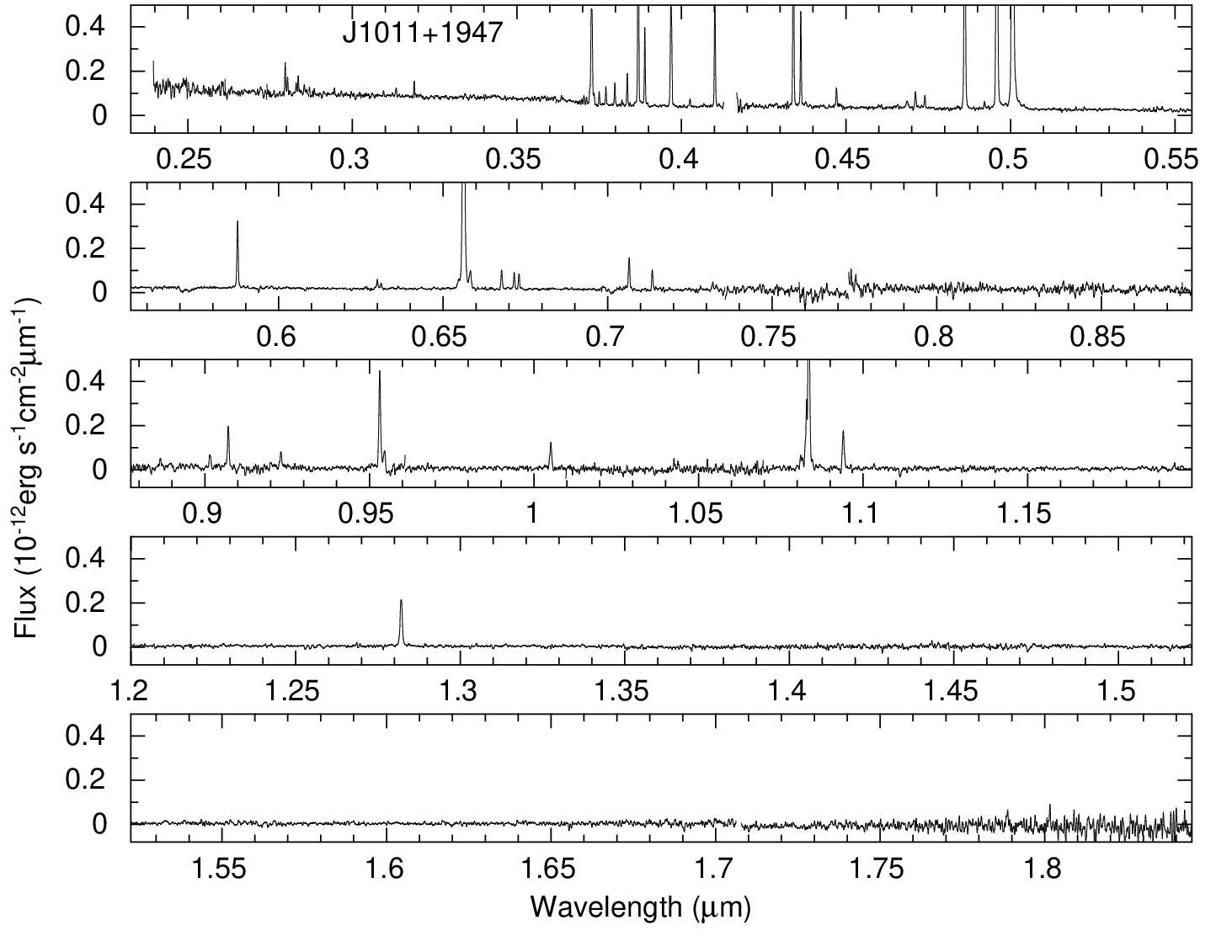}
\includegraphics[angle=-90,width=0.85\linewidth]{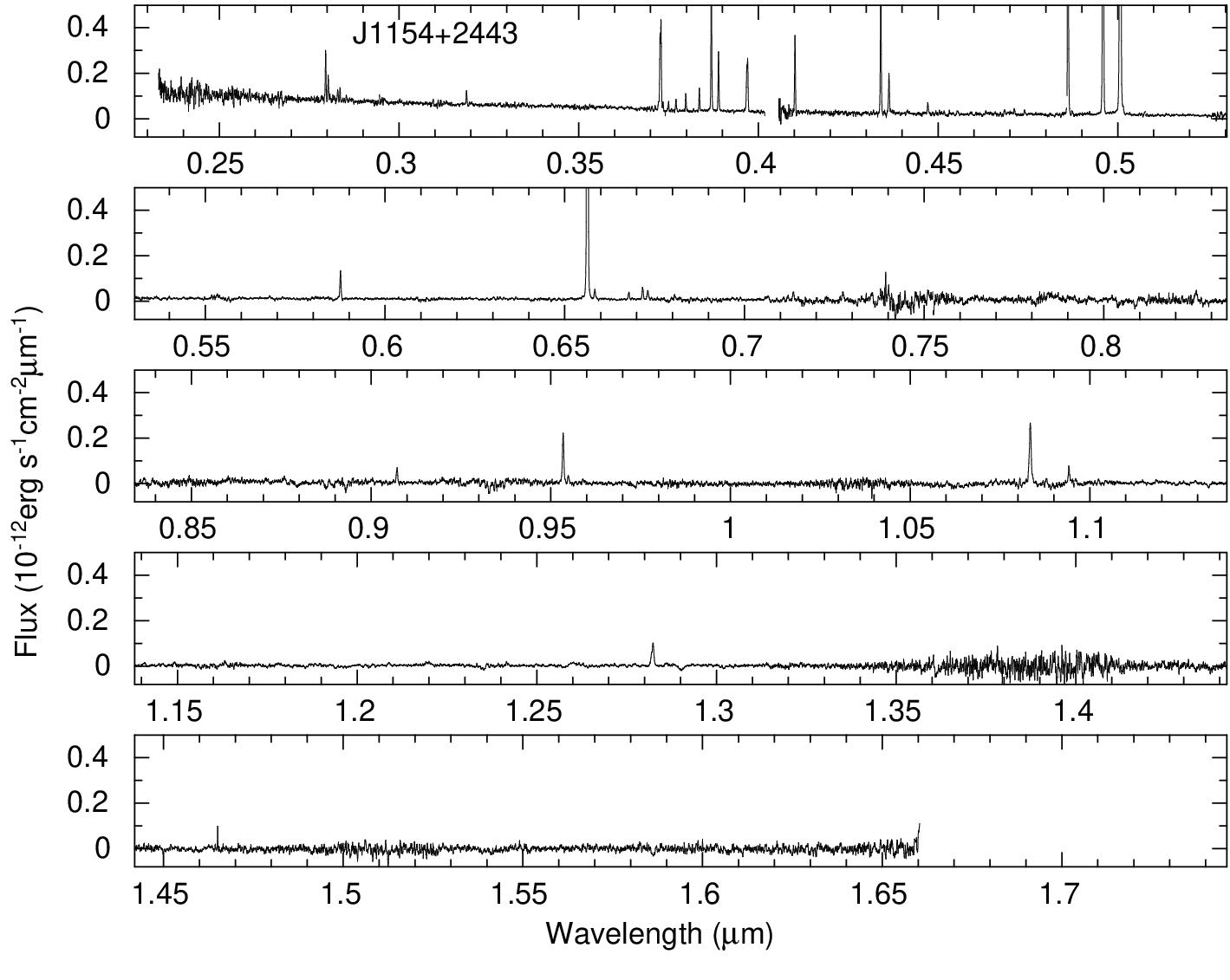}
\caption{Continued.}
\end{figure*}

\setcounter{figure}{0}

\begin{figure*}
\includegraphics[angle=-90,width=0.85\linewidth]{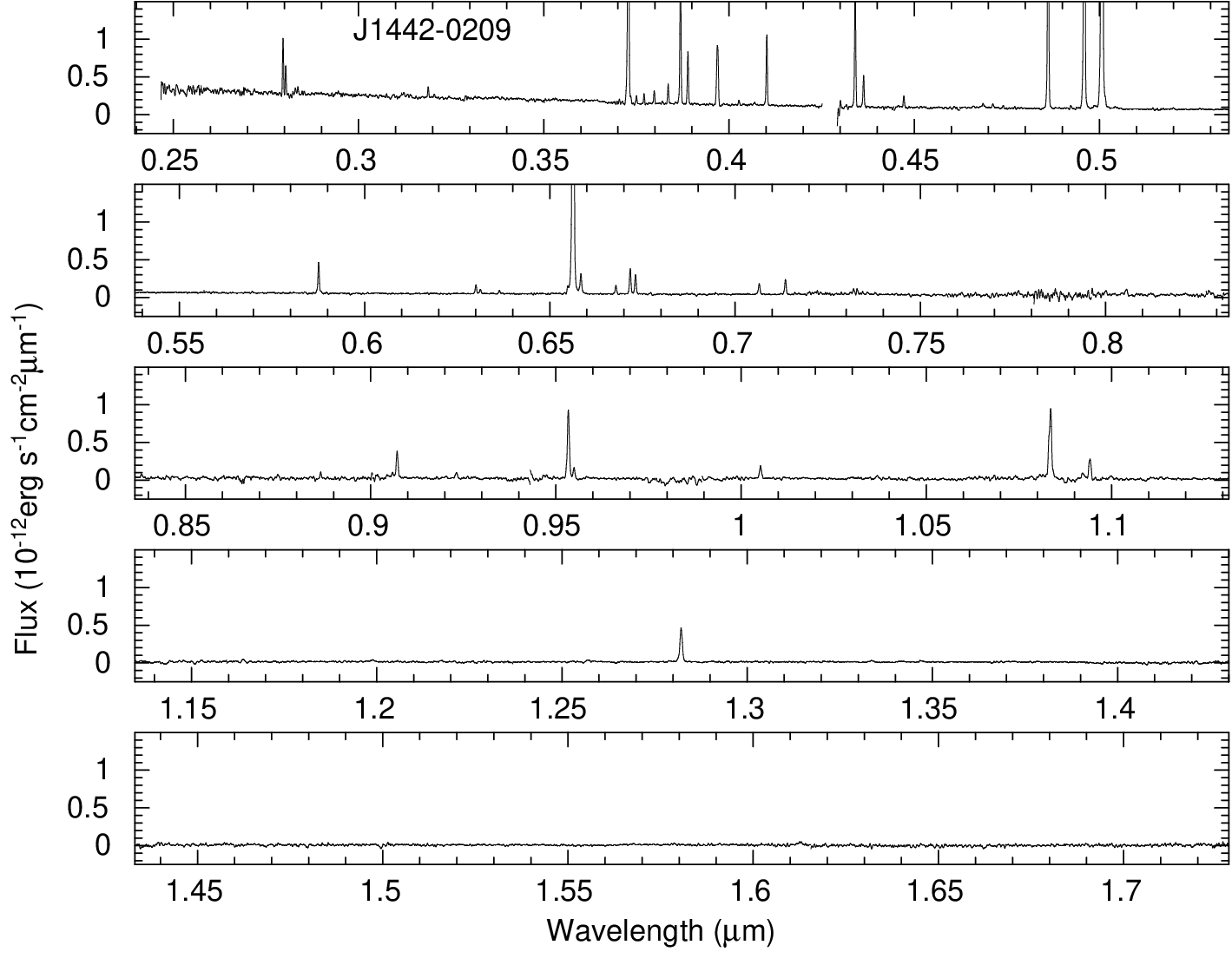}
\caption{Continued.}
\label{fig1}
\end{figure*}

\begin{landscape}
\begin{table} 
 \caption{Emission Line Intensities and Equivalent Widths \label{tab1}}
 \begin{tabular}{lrrrrrrrrrr} \hline \hline
&\multicolumn{10}{c}{Galaxy}\\  \hline      
 &\multicolumn{2}{c}{          J0901+2119}&
 \multicolumn{2}{c}{          J0925+1403}&
 \multicolumn{2}{c}{          J1011+1947}&
 \multicolumn{2}{c}{          J1154+2443}&
 \multicolumn{2}{c}{          J1442-0209} \\ 
  {Ion}
  &\multicolumn{1}{c}{$I$($\lambda$)/$I$(H$\beta$)$^{\rm a}$}
  &\multicolumn{1}{c}{EW$^{\rm b}$}
  &\multicolumn{1}{c}{$I$($\lambda$)/$I$(H$\beta$)$^{\rm a}$}
  &\multicolumn{1}{c}{EW$^{\rm b}$}
  &\multicolumn{1}{c}{$I$($\lambda$)/$I$(H$\beta$)$^{\rm a}$}
  &\multicolumn{1}{c}{EW$^{\rm b}$}
  &\multicolumn{1}{c}{$I$($\lambda$)/$I$(H$\beta$)$^{\rm a}$}
  &\multicolumn{1}{c}{EW$^{\rm b}$}
  &\multicolumn{1}{c}{$I$($\lambda$)/$I$(H$\beta$)$^{\rm a}$}
  &\multicolumn{1}{c}{EW$^{\rm b}$} \\ \hline
 2613 Fe {\sc ii}*                &   1.18 $\pm$   0.27 &   0.8 & 1.11 $\pm$   0.21 &  0.4&   ...~~~~~~ &  ...~~&   ...~~~~~~ &  ...~~&   1.90 $\pm$   0.49 &   0.7 \\
 2626 Fe {\sc ii}*                &   0.74 $\pm$   0.19 &   0.5 &  1.56 $\pm$ 0.22&    0.5&   ...~~~~~~ &  ...~~&   ...~~~~~~ &  ...~~&   2.17 $\pm$   0.10 &   0.8 \\
 2796 Mg {\sc ii}                 &  14.88 $\pm$   0.82 &  10.7 &  14.35 $\pm$   0.86 &   6.1 &   6.08 $\pm$   0.48 &   5.0 &  12.48 $\pm$   0.97 &   6.1 &  24.57 $\pm$   1.33 &  11.9 \\
 2803 Mg {\sc ii}                 &   7.61 $\pm$   0.61 &   5.7 &   7.15 $\pm$   0.70 &   3.1 &   3.37 $\pm$   0.38 &   2.8 &   6.66 $\pm$   0.89 &   3.3 &  12.91 $\pm$   1.01 &   6.4 \\
 3188 He {\sc i}                  &   3.28 $\pm$   0.19 &   3.0 &   3.31 $\pm$   0.27 &   1.9 &   2.54 $\pm$   0.35 &   2.4 &   3.29 $\pm$   0.31 &   2.0 &   3.67 $\pm$   0.29 &   2.1 \\
 3322 [Fe {\sc ii}]               &   0.37 $\pm$   0.10 &   0.4 &   ...~~~~~~ &  ...~~&   ...~~~~~~ &  ...~~&   ...~~~~~~ &  ...~~&   ...~~~~~~ &  ...~~\\
 3692 H18                         &   0.50 $\pm$   0.09 &   0.8 &   0.69 $\pm$   0.15 &   0.6 &   0.79 $\pm$   0.12 &   1.5 &   0.98 $\pm$   0.15 &   1.1 &   ...~~~~~~ &  ...~~\\
 3697 H17                         &   0.50 $\pm$   0.10 &   0.7 &   1.17 $\pm$   0.22 &   1.1 &   1.07 $\pm$   0.11 &   2.0 &   1.29 $\pm$   0.15 &   1.5 &   1.54 $\pm$   0.27 &   1.4 \\
 3703 H16+O {\sc iii}             &   1.50 $\pm$   0.15 &   2.6 &   1.96 $\pm$   0.20 &   1.9 &   1.82 $\pm$   0.14 &   3.4 &   1.32 $\pm$   0.20 &   1.5 &   1.76 $\pm$   0.31 &   1.7 \\
 3712 H15                         &   3.54 $\pm$   0.41 &   2.5 &   4.19 $\pm$   0.62 &   1.2 &   3.08 $\pm$   0.32 &   2.8 &   4.79 $\pm$   0.64 &   2.2 &   3.65 $\pm$   0.82 &   1.4 \\
 3722 H14                         &   1.31 $\pm$   0.10 &   0.9 &   0.48 $\pm$   0.08 &   0.2 &  14.98 $\pm$   0.50 &  29.7 &   2.90 $\pm$   0.25 &   2.8 &   1.79 $\pm$   0.06 &   1.0 \\
 3727 [O {\sc ii}]                &  90.12 $\pm$   2.88 & 192.8 & 109.81 $\pm$   3.53 & 121.8 &  28.66 $\pm$   0.94 &  57.9 &  49.33 $\pm$   1.67 &  56.4 &  94.27 $\pm$   3.04 &  92.7 \\
 3734 H13                         &   1.83 $\pm$   0.12 &   2.8 &   2.22 $\pm$   0.18 &   2.0 &   2.18 $\pm$   0.21 &   3.5 &   2.47 $\pm$   0.24 &   2.8 &   2.84 $\pm$   0.25 &   2.8 \\
 3750 H12                         &   5.18 $\pm$   0.26 &   7.6 &   5.20 $\pm$   0.45 &   2.5 &   4.06 $\pm$   0.26 &   5.6 &   5.29 $\pm$   0.75 &   3.0 &   5.66 $\pm$   0.45 &   3.3 \\
 3771 H11                         &   5.97 $\pm$   0.28 &   8.1 &   6.31 $\pm$   0.38 &   3.4 &   5.05 $\pm$   0.30 &   6.4 &   6.91 $\pm$   0.61 &   4.8 &   6.20 $\pm$   0.49 &   3.9 \\
 3798 H10                         &   6.66 $\pm$   0.28 &   8.4 &   7.74 $\pm$   0.37 &   5.4 &   5.47 $\pm$   0.30 &   7.4 &   7.94 $\pm$   0.59 &   5.8 &   7.14 $\pm$   0.43 &   4.3 \\
 3820 He {\sc i}                  &   1.40 $\pm$   0.09 &   2.7 &   0.71 $\pm$   0.10 &   0.7 &   1.22 $\pm$   0.11 &   2.8 &   ...~~~~~~ &  ...~~&   0.93 $\pm$  0.09 &   0.9 \\
 3835 H9                          &   9.18 $\pm$   0.34 &  14.5 &   9.43 $\pm$   0.41 &   6.6 &   7.39 $\pm$   0.36 &  13.2 &  10.25 $\pm$   0.53 &   7.9 &   9.87 $\pm$   0.45 &   7.1 \\
 3869 [Ne {\sc iii}]              &  55.21 $\pm$   1.74 &  98.7 &  49.90 $\pm$   1.60 &  50.2 &  48.92 $\pm$   1.58 &  83.3 &  42.18 $\pm$   1.38 &  44.5 &  53.81 $\pm$   1.72 &  52.5 \\
 3889 He {\sc i} + H8             &  20.54 $\pm$   0.67 &  38.7 &  21.59 $\pm$   0.74 &  19.0 &  15.42 $\pm$   0.56 &  23.2 &  22.24 $\pm$   0.80 &  22.0 &  21.83 $\pm$   0.77 &  17.7 \\
 3968 [Ne {\sc iii}] + H7         &  17.55 $\pm$   0.58 &  36.8 &  14.59 $\pm$   0.51 &  13.2 &  13.55 $\pm$   0.45 &  34.6 &  16.92 $\pm$   0.65 &  18.2 &  18.15 $\pm$   0.65 &  19.0 \\
 4026 He {\sc i}                  &   1.46 $\pm$   0.08 &   2.8 &   1.66 $\pm$   0.13 &   1.8 &   1.38 $\pm$   0.12 &   3.1 &   ...~~~~~~ &  ...~~&   2.43 $\pm$   0.23 &   2.9 \\
 4069 [S {\sc ii}]                &   1.10 $\pm$   0.07 &   2.2 &   1.03 $\pm$   0.10 &   1.2 &   0.46 $\pm$   0.06 &   1.1 &   ...~~~~~~ &  ...~~&   1.11 $\pm$   0.17 &   1.2 \\
 4076 [S {\sc ii}]                &   0.50 $\pm$   0.05 &   1.1 &   0.53 $\pm$   0.07 &   0.6 &   0.20 $\pm$   0.03 &   0.5 &   ...~~~~~~ &  ...~~&   0.45 $\pm$   0.53 &   0.5 \\
 4101 H$\delta$                   &  28.53 $\pm$   0.88 &  61.2 &  28.64 $\pm$   0.91 &  30.2 &  24.00 $\pm$   0.80 &  47.5 &  26.21 $\pm$   1.13 &  37.4 &  27.93 $\pm$   0.91 &  29.0 \\
 4340 H$\gamma$                   &  46.06 $\pm$   1.40 & 103.8 &  46.69 $\pm$   1.44 &  63.5 &  47.90 $\pm$   1.49 & 178.6 &  45.56 $\pm$   1.58 &  64.2 &  47.01 $\pm$   1.49 &  61.4 \\
 4363 [O {\sc iii}]               &  10.81 $\pm$   0.37 &  26.3 &   7.23 $\pm$   0.31 &  11.1 &  16.38 $\pm$   0.56 &  35.4 &  13.43 $\pm$   0.76 &  24.5 &  10.63 $\pm$   0.43 &  14.8 \\
 4471 He {\sc i}                  &   4.08 $\pm$   0.23 &  12.7 &   4.07 $\pm$   0.24 &   6.4 &   3.27 $\pm$   0.31 &   8.4 &   3.95 $\pm$   0.52 &   7.9 &   3.56 $\pm$   0.31 &   5.2 \\
 4658 [Fe {\sc iii}]              &   0.73 $\pm$   0.13 &   2.2 &   1.12 $\pm$   0.13 &   1.9 &   ...~~~~~~ &  ...~~&   ...~~~~~~ &  ...~~&   ...~~~~~~ &  ...~~\\
 4686 He {\sc ii}                 &   0.88 $\pm$   0.18 &   2.7 &   1.13 $\pm$   0.16 &   1.9 &   1.78 $\pm$   0.25 &   5.3 &   2.02 $\pm$   0.45 &   4.7 &   0.99 $\pm$   0.13 &   1.4 \\
 4713 [Ar {\sc iv}] + He {\sc i}  &   2.52 $\pm$   0.19 &   9.0 &   1.49 $\pm$   0.26 &   2.8 &   2.91 $\pm$   0.16 &   7.5 &   2.66 $\pm$   0.35 &   6.6 &   1.03 $\pm$   0.16 &   1.6 \\
 4740 [Ar {\sc iv}]               &   1.14 $\pm$   0.24 &   3.9 &   0.86 $\pm$   0.20 &   1.6 &   2.27 $\pm$   0.18 &   6.6 &   2.82 $\pm$   0.37 &   9.5 &   0.77 $\pm$   0.16 &   1.4 \\
 4861 H$\beta$                    & 100.00 $\pm$   2.88 & 357.5 & 100.00 $\pm$   2.91 & 208.1 & 100.00 $\pm$   2.89 & 314.9 & 100.00 $\pm$   2.99 & 255.1 & 100.00 $\pm$   2.91 & 186.8 \\
 4922 He {\sc i}                  &   0.89 $\pm$   0.14 &   3.1 &   0.76 $\pm$   0.12 &   1.5 &   1.42 $\pm$   0.11 &   4.8 &   ...~~~~~~ &  ...~~&   0.99 $\pm$   0.15 &   1.9 \\
 4959 [O {\sc iii}]               & 232.94 $\pm$   6.70 & 833.9 & 198.62 $\pm$   5.76 & 387.4 & 278.38 $\pm$   8.02 & 774.7 & 188.89 $\pm$   5.55 & 356.6 & 209.38 $\pm$   6.08 & 358.1 \\
 4986 [Fe {\sc iii}]              &   0.83 $\pm$   0.10 &   2.6 &   0.54 $\pm$   0.07 &   1.2 &   ...~~~~~~ &  ...~~&   ...~~~~~~ &  ...~~&   ...~~~~~~ &  ...~~\\
 5007 [O {\sc iii}]               & 694.26 $\pm$  19.96 &2229.0 & 592.10 $\pm$  17.13 &1317.0 & 826.75 $\pm$  23.83 &1848.0 & 573.15 $\pm$  17.05 & 720.5 & 643.46 $\pm$  18.61 &1097.0 \\
 5016 He {\sc i}                  &   1.18 $\pm$   0.06 &   2.4 &   1.28 $\pm$   0.11 &   1.8 &   ...~~~~~~ &  ...~~&   ...~~~~~~ &  ...~~&   1.34 $\pm$   0.10 &   1.3 \\
 5041 Si {\sc ii}                 &   0.27 $\pm$   0.08 &   1.0 &   ...~~~~~~ &  ...~~&   ...~~~~~~ &  ...~~&   ...~~~~~~ &  ...~~&   0.53 $\pm$   0.08 &   1.0 \\
 5048 He {\sc i}                  &   0.27 $\pm$   0.07 &   1.1 &   ...~~~~~~ &  ...~~&   ...~~~~~~ &  ...~~&   ...~~~~~~ &  ...~~&   ...~~~~~~ &  ...~~\\
 5056 Si {\sc ii}                 &   0.46 $\pm$   0.07 &   1.9 &   ...~~~~~~ &  ...~~&   ...~~~~~~ &  ...~~&   ...~~~~~~ &  ...~~&   ...~~~~~~ &  ...~~\\
 5755 [N {\sc ii}]                &   0.43 $\pm$   0.08 &   2.4 &   ...~~~~~~ &  ...~~&   ...~~~~~~ &  ...~~&   ...~~~~~~ &  ...~~&   ...~~~~~~ &  ...~~\\
 \hline
  \end{tabular}
  \end{table}
\end{landscape}  

\begin{landscape}
\begin{table} 
 \contcaption{}
  \begin{tabular}{lrrrrrrrrrr}  \hline \hline
&\multicolumn{10}{c}{Galaxy}\\  \hline      
 &\multicolumn{2}{c}{          J0901+2119}&
 \multicolumn{2}{c}{          J0925+1403}&
 \multicolumn{2}{c}{          J1011+1947}&
 \multicolumn{2}{c}{          J1154+2443}&
 \multicolumn{2}{c}{          J1442-0209} \\ 
  {Ion}
  &\multicolumn{1}{c}{$I$($\lambda$)/$I$(H$\beta$)$^{\rm a}$}
  &\multicolumn{1}{c}{EW$^{\rm b}$}
  &\multicolumn{1}{c}{$I$($\lambda$)/$I$(H$\beta$)$^{\rm a}$}
  &\multicolumn{1}{c}{EW$^{\rm b}$}
  &\multicolumn{1}{c}{$I$($\lambda$)/$I$(H$\beta$)$^{\rm a}$}
  &\multicolumn{1}{c}{EW$^{\rm b}$}
  &\multicolumn{1}{c}{$I$($\lambda$)/$I$(H$\beta$)$^{\rm a}$}
  &\multicolumn{1}{c}{EW$^{\rm b}$}
  &\multicolumn{1}{c}{$I$($\lambda$)/$I$(H$\beta$)$^{\rm a}$}
  &\multicolumn{1}{c}{EW$^{\rm b}$} \\ \hline
 5876 He {\sc i}                  &  11.25 $\pm$   0.38 &  55.9 &  10.91 $\pm$   0.36 &  33.1 &  12.39 $\pm$   0.39 &  60.5 &  11.46 $\pm$   0.43 &  53.0 &  10.24 $\pm$   0.73 &  35.7 \\
 6300 [O {\sc i}]                 &   2.65 $\pm$   0.11 &  15.6 &   3.07 $\pm$   0.12 &  11.9 &   1.73 $\pm$   0.14 &  11.1 &   2.13 $\pm$   0.20 &  10.2 &   2.99 $\pm$   0.13 &  11.0 \\
 6312 [S {\sc iii}]               &   1.31 $\pm$   0.08 &   8.2 &   1.27 $\pm$   0.09 &   5.1 &   0.97 $\pm$   0.11 &   6.7 &   1.12 $\pm$   0.17 &   5.1 &   1.29 $\pm$   0.08 &   4.5 \\
 6364 [O {\sc i}]                 &   0.94 $\pm$   0.06 &   5.5 &   1.02 $\pm$   0.09 &   4.1 &   0.59 $\pm$   0.08 &   5.0 &   1.09 $\pm$   0.20 &   5.8 &   1.00 $\pm$   0.09 &   3.7 \\
 6548 [N {\sc ii}]                &   3.12 $\pm$   0.12 &  15.1 &   2.45 $\pm$   0.12 &   4.6 &   1.54 $\pm$   0.14 &   8.4 &   1.77 $\pm$   0.31 &   8.9 &   1.66 $\pm$   0.10 &   4.0 \\
 6563 H$\alpha$                   & 283.65 $\pm$   8.82 &1902.0 & 285.04 $\pm$   8.92 &1105.0 & 281.09 $\pm$   8.77 &2503.0 & 277.61 $\pm$   8.82 &1361.0 & 282.85 $\pm$   8.85 &1086.0 \\
 6583 [N {\sc ii}]                &  11.83 $\pm$   0.38 &  75.2 &  11.77 $\pm$   0.38 &  46.8 &   3.95 $\pm$   0.18 &  21.5 &   4.90 $\pm$   0.27 &  22.1 &   8.42 $\pm$   0.29 &  31.9 \\
 6678 He {\sc i}                  &   3.43 $\pm$   0.13 &  28.4 &   3.08 $\pm$   0.12 &  13.5 &   3.35 $\pm$   0.13 &  25.5 &   3.46 $\pm$   0.20 &  19.2 &   2.93 $\pm$   0.12 &  13.6 \\
 6716 [S {\sc ii}]                &   7.74 $\pm$   0.26 &  53.0 &   9.40 $\pm$   0.31 &  43.9 &   2.39 $\pm$   0.10 &  15.5 &   4.57 $\pm$   0.24 &  19.5 &   8.68 $\pm$   0.30 &  34.2 \\
 6731 [S {\sc ii}]                &   6.90 $\pm$   0.23 &  40.5 &   7.77 $\pm$   0.27 &  34.3 &   2.36 $\pm$   0.10 &  16.5 &   3.64 $\pm$   0.27 &  14.3 &   6.51 $\pm$   0.23 &  25.6 \\
 7065 He {\sc i}                  &   5.01 $\pm$   0.18 &  36.2 &   4.15 $\pm$   0.17 &  18.8 &   5.58 $\pm$   0.22 &  36.3 &   4.45 $\pm$   0.44 &  35.5 &   4.14 $\pm$   0.17 &  20.2 \\
 7136 [Ar {\sc iii}]              &   5.92 $\pm$   0.23 &  47.9 &   5.67 $\pm$   0.23 &  28.0 &   3.40 $\pm$   0.15 &  36.3 &   3.53 $\pm$   0.48 &  19.3 &   5.02 $\pm$   0.21 &  22.9 \\
 7320 [O {\sc ii}]                &   2.02 $\pm$   0.11 &  16.6 &   2.11 $\pm$   0.13 &  11.8 &   ...~~~~~~ &  ...~~&   ...~~~~~~ &  ...~~&   1.53 $\pm$   0.10 &   5.8 \\
 7330 [O {\sc ii}]                &   1.41 $\pm$   0.12 &  11.7 &   1.75 $\pm$   0.13 &   9.8 &   ...~~~~~~ &  ...~~&   ...~~~~~~ &  ...~~&   1.51 $\pm$   0.14 &   6.0 \\
 8750 P12                         &   ...~~~~~~ &  ...~~&   ...~~~~~~ &  ...~~&   ...~~~~~~ &  ...~~&   ...~~~~~~ &  ...~~&   1.71 $\pm$   0.24 &  17.1 \\
 8863 P11                         &   ...~~~~~~ &  ...~~&   ...~~~~~~ &  ...~~&   ...~~~~~~ &  ...~~&   ...~~~~~~ &  ...~~&   1.64 $\pm$   0.18 &  15.0 \\
 9015 P10                         &   ...~~~~~~ &  ...~~&   1.27 $\pm$   0.12 &  26.6 &   ...~~~~~~ &  ...~~&   ...~~~~~~ &  ...~~&   ...~~~~~~ &  ...~~\\
 9069 [S {\sc iii}]               &   7.25 $\pm$   0.31 & 210.0 &   9.06 $\pm$   0.39 & 205.4 &   8.24 $\pm$   0.42 &  68.6 &  10.99 $\pm$   1.11 & 146.7 &   9.47 $\pm$   0.49 &  36.7 \\
 9229 P9                          &   1.95 $\pm$   0.12 &  35.4 &   1.35 $\pm$   0.11 &  19.1 &   2.62 $\pm$   0.22 &  19.1 &   ...~~~~~~ &  ...~~&   2.15 $\pm$   0.20 &  14.8 \\
 9531 [S {\sc iii}]               &  16.85 $\pm$   0.65 & 797.7 &  19.91 $\pm$   0.78 & 370.1 &  20.48 $\pm$   0.84 & 223.3 &  32.57 $\pm$   1.52 & 169.6 &  28.87 $\pm$   1.16 & 221.8 \\
 9546 P8                          &   2.05 $\pm$   0.11 &  98.2 &   1.76 $\pm$   0.10 &  24.9 &   4.02 $\pm$   0.24 &  60.8 &   6.88 $\pm$   0.66 & 140.6 &   4.06 $\pm$   0.23 &  37.0 \\
 10049 P7                         &   3.08 $\pm$   0.14 & 124.6 &   2.95 $\pm$   0.15 &  62.7 &   5.56 $\pm$   0.30 &  90.2 &   ...~~~~~~ &  ...~~&   5.10 $\pm$   0.28 &  41.3 \\ 
 10829 He {\sc i}                 &  14.42 $\pm$   0.60 & 984.7 &   5.21 $\pm$   0.29 & 167.0 &  54.84 $\pm$   2.20 &1051.0 &  55.27 $\pm$   2.50 & 122.3 &   36.23 $\pm$  1.53 &  119.1 \\
 10941 P$\gamma$                  &   3.88 $\pm$   0.21 & 193.9 &   0.98 $\pm$   0.10 &  23.9 &   8.54 $\pm$   0.40 & 239.5 &  13.98 $\pm$   1.56 & 828.0 &   9.14 $\pm$   0.48 & 216.5 \\
 12570 [Fe {\sc ii}]              &   0.47 $\pm$   0.03 &  24.6 &   ...~~~~~~ &  ...~~&   ...~~~~~~ &  ...~~&   ...~~~~~~ &  ...~~&   1.25 $\pm$   0.10 &  21.2 \\
 12790 He {\sc i}                 &   0.44 $\pm$   0.04 &  32.3 &   0.41 $\pm$   0.05 &  14.6 &   ...~~~~~~ &  ...~~&   ...~~~~~~ &  ...~~&   ...~~~~~~ &  ...~~\\
 12821 P$\beta$                   &   7.12 $\pm$   0.30 & 432.5 &   7.56 $\pm$   0.32 & 282.1 &  13.28 $\pm$   0.56 & 455.1 &  16.42 $\pm$   0.96 & 319.4 &  17.13 $\pm$   0.73 & 257.8 \\
 16430 [Fe {\sc ii}]              &   0.25 $\pm$   0.02 &  61.8 &   ...~~~~~~ &  ...~~&   ...~~~~~~ &  ...~~&   ...~~~~~~ &  ...~~&   ...~~~~~~ &  ...~~ \\  
 18756 P$\alpha$                  &  10.64 $\pm$   0.58 &1816.0 &  20.96 $\pm$   1.03 & 347.0 &   ...~~~~~~ &  ...~~&   ...~~~~~~ &  ...~~&   ...~~~~~~ &  ...~~\\
\\
EW(abs)$^{\rm c}$  & \multicolumn{2}{c}{ 3.59 }& \multicolumn{2}{c}{ 2.81 }& \multicolumn{2}{c}{ 3.30 }& \multicolumn{2}{c}{ 3.39 }& \multicolumn{2}{c}{ 2.47 } \\
 $C$(H$\beta$)$^{\rm d}$ & \multicolumn{2}{c}{ 0.180 }& \multicolumn{2}{c}{ 0.145 }& \multicolumn{2}{c}{ 0.155 }& \multicolumn{2}{c}{ 0.020 }& \multicolumn{2}{c}{ 0.170 } \\
 $F$(H$\beta$)$^{\rm e}$ & \multicolumn{2}{c}{  11.78 }& \multicolumn{2}{c}{ 21.61 }& \multicolumn{2}{c}{ 13.12 }& \multicolumn{2}{c}{  9.77 }& \multicolumn{2}{c}{ 19.44 } \\
\hline  \hline
\end{tabular}
 
 \hbox{$^{\rm a}$Ratio of extinction corrected fluxes to H$\beta$ multipled by 100.}

 \hbox{$^{\rm b}$Equivalent width of emission lines in \AA.}

\hbox{$^{\rm c}$Equivalent widths of underlying hydrogen absorption lines in \AA.}

\hbox{$^{\rm d}$Extinction coefficient, obtained from Balmer decrement.} 

 \hbox{$^{\rm e}$Observed flux in units of 10$^{-16}$ erg s$^{-1}$ cm$^{-2}$.}

  \end{table}
\end{landscape}

\bsp

\label{lastpage}


\begin{thebibliography}{}

\bibitem[Amorin et al.(2010)Amorin, P\'erez-Montero \& V\'ilchez]{Amorin2010} 
Amorin R. O., P\'erez-Montero E., V\'ilchez J. M., 2010, \apjl, 715, L128

\bibitem[Amorin et al.(2012)]{Amorin2012a} Amorin R. O., V\'ilchez J. M.,
  P\'erez-Montero E., Papaderos P., 2012, \apj, 749, 185

\bibitem[Amorin et al.(2012)]{Amorin2012} Amorin R. O., V\'ilchez J. M.,
  P\'erez-Montero E., 2012, \assp, 28, 243 

\bibitem[Andrews \&  Martini(2013)]{AndrewsMartini2013} Andrews B. H.,
  Martini P., 2013, \apj, 765, 140

\bibitem[Baldwin et al.(1981)Baldwin, Phillips \& Terlevich]{BPT81} 
Baldwin J. A., Phillips M. M., Terlevich R., 1981, \pasp, 93, 5

\bibitem[Bassett et al.(2019)]{Bassett2019} Bassett R., Ryan-Weber E. V.,
Cooke J. et al., 2019, \mnras, 483, 5223

\bibitem[Bayliss et al.(2014)]{Bayliss2014} Bayliss M. B., Rigby J. R.,
  Sharon K., Wuyts E., Florian M., Gladders M. D., Johnson T., Oguri, M., 
2014, \apj, 790, 144

\bibitem[Becker et al.(2015)]{Becker2015} Becker G. D., Bolton J. S.,
Lidz A., 2015, \pasa, 32, 45

\bibitem[Bian et al.(2020)]{Bian2020} Bian F., Kewley L. J., Groves B.,
 Dopita M. A., 2020, \mnras, 493, 580

\bibitem[Bian \& Fan(2020a)]{Bian2020a} Bian F., Fan X., 2020a, \mnras,
  493, L65
 
\bibitem[Bowler et al.(2017)]{Bowler2017} Bowler R. A. A., Dunlop J. S.,
  McLure R. J., McLeod D. J., 2017, \mnras, 466, 3612

\bibitem[Bouwens et al.(2004)]{Bouwens2004} Bouwens R. J.,
Illingworth G. D., Blakeslee J. P., Broadhurst T. J., Franx M., 2004a,
\apjl, 611, L1

\bibitem[Bouwens et al.(2016)]{Bouwens2016} Bouwens R. J., Aravena M.,
Decarli R. et al., 2016, \apj, 833, 72

\bibitem[Castellano et al.(2017)]{Castellano2017} Castellano M.,
 Pentericci L., Fontana A. et al.,
  2017, \apj, 839, 73

\bibitem[Cardelli et al.(1989)Cardelli, Clayton \& Mathis]{Cardelli1989} Cardelli J. A., Clayton G. C.,
  Mathis J. S., 1989, \apj, 345, 245

\bibitem[Charlot \& Longhetti (2001)]{Charlot2001} Charlot S., Longhetti M.,
 2001, \mnras, 323, 887

\bibitem[Chevallard et al.(2018)]{Chevallard2018} Chevallard J., Charlot S.,
Senchyna P. et al., 2018, \mnras, 479, 3264

\bibitem[Christensen et al.(2012)]{Christensen2012} Christensen L., Laursen P.,
Richard J. et al, 2012, \mnras, 427, 1973

\bibitem[Curtis-Lake et al.(2016)]{CurtisLake2016} Curtis-Lake E.,
McLure R. J., Dunlop J. S. et al., 2016, \mnras, 457, 440

\bibitem[de Barros et al.(2014)]{deBarros2014} de Barros S., Schaerer D.,
  Stark D. P., 2014, \aap, 563, A81

\bibitem[de Barros et al.(2016)]{deBarros2016} de Barros S., Vanzella E.,
Amor\'in R. et al.,  2016, \aap, 585, A51  

\bibitem[Dors et al.(2018)]{Dors2018} Dors O. L., Agarwal B.,
  H\"{a}gele G. F., Cardaci M. V., Rydberg C.-E., Riffel R. A., Oliveira A. S.,
  Krabbe A. C., 2018, \mnras, 479, 2294

\bibitem[Duncan et al.(2014)]{Duncan2014} Duncan K., Conselice C. J.,
Mortlock A. et al., 2014, \mnras, 444, 2960

\bibitem[Eldridge et al.(2017)]{Eldridge2017} Eldridge J. J., Stanway E. R.,
  Xiao L., McClelland L. A. S., Taylor G., Ng M., Greis S. M. L., Bray J. C.,
2017, \pasa, 34, 58

\bibitem[Emami et al.(2019)]{Emami2019} Emami N., Siana B., Alavi A., 
Gburek T., Freeman W. R., Richard J., Weisz D. R., Stark D. P.,
  2019, arXiv:1912.06152 

\bibitem[Endsley et al.(2020)]{Endsley2020} Endsley R., Stark D. P.,
Chevallard J., Charlot S., 2020, arXiv:2005.02402 
  
\bibitem[Erb et al.(2012)]{Erb2012} Erb D. K., Quider A. M., Henry A. L.,
  Martin C. L., 2012, \apj, 759, 26

\bibitem[Erb et al.(2016)]{Erb2016} Erb D. K., Pettini M., Steidel C. C.,
 Strom A. L., Rudie G. C., Trainor R. F., Shapley A. E., Reddy N. A., 
2016, \apj, 830, 52

\bibitem[Faisst et al.(2019)]{Faisst2019} Faisst A. L., Capak P. L., Emami N.,
Tacchella S., Larson K. L., 2019, \apj, 884, 133

\bibitem[Feltre et al.(2018)]{Feltre2018} Feltre A., Bacon R., Finley H.
  et al., 2018, \aap, 617, A62

\bibitem[Ferguson et al.(2004)]{Ferguson2004} Ferguson H. C.,
Dickinson M., Giavalisco M. et al., 2004, \apjl, 600, L107

\bibitem[Ferland et al.(2017)]{Ferland17} Ferland G. J., Chatzikos M., 
Guzm\'an F. et al., 2017, \rmxaa, 53, 385

\bibitem[Finley et al.(2017)]{Finley2017} Finley H., Bouch\'e N.,
  Contini T. et al., 2017, \aap, 608, A7

\bibitem[Fischer \& Tachiev(2014)]{Fische2014} Fischer C. F., Tachiev G.,
2014, MCHF/MCDHF Collection, Version 2, Ref No. 10 \& 20, National Institute of
Standards and Technology, http://physics.nist.gov/mchf

\bibitem[Fletcher et al.(2019)]{Fletcher2019} Fletcher T. J., Tang M.,
  Robertson B. E., Nakajima K., Ellis R. S., 
Stark D. P., Inoue A., 2019, \apj, 878, 87

\bibitem[Gazagnes et al.(2020)]{Gazagnes2020} Gazagnes S., Chisholm J.,
  Schaerer D., Verhamme A., Izotov Y., 2020, \aap, 639, 85

 \bibitem[Gonz\'{a}lez et al.(2014)]{Gonzalez2014} Gonz\'{a}lez V.,
    Bouwens R., Illingworth G.,  Labb\'{e} I., 
Oesch p., Franx M., Magee D., 2014, \apj, 781, 34

\bibitem[Gonz\'alez-Delgado, Leitherer \& Heckman(1999)]{Gonzalez1999}
  Gonz\'alez-Delgado R. M., Leitherer C., Heckman T. M., 1999, \apjs,
  125, 489
  
\bibitem[G\"otberg et al.(2018)]{Gotberg2018} G\"otberg Y.,  de Mink S. E.,
  Groh J. H., Kupfer T., Crowther P. A., Zapartas E., Renzo M., 
2018, \aap, 615, 78
  
\bibitem[Grazian et al.(2015)]{Grazian2015} Grazian A., Fontana A.,
Santini P. et al., 2015, \aap, 575, 96

\bibitem[Guseva et al.(2000)Guseva, Izotov \& Thuan]{GIT2000} Guseva N. G., 
Izotov Y. I., Thuan T. X., 2000, \apj, 531, 776

\bibitem[Guseva et al.(2011)Guseva, Izotov \& Stasi\'nska]{Guseva2011}
Guseva N. G., Izotov Y. I., Stasi\'nska G., 2011, \aap, 529, 149 
   
\bibitem[Guseva et al.(2012)]{Guseva2012} Guseva N. G., Izotov Y. I.,
Fricke K. J., Henkel C., 2012, \aap, 541, 115

\bibitem[Guseva et al.(2013)]{Guseva2013} Guseva N. G., Izotov Y. I.,
Fricke K. J., Henkel C., 2013, \aap, 555, A90

\bibitem[Guseva et al.(2015)]{Guseva2015} Guseva N. G., Izotov Y. I.,
Fricke K. J., Henkel C., 2015, \aap, 579, 11

\bibitem[Guseva et al.(2019)]{Guseva2019} Guseva N. G., Izotov Y. I.,
Fricke K. J., Henkel C., 2019, \aap, 624, A21

\bibitem[Harshan et al.(2020)]{Harshan2020} Harshan A., Gupta A., Tran K.-V.
  et al., 2020, arXiv:2002.08353

\bibitem[Hayashi et al.(2015)]{Hayashi2015} Hayashi M., Ly C.,
Shimasaku K. et al., 2015, \pasj, 67, 80

\bibitem[Henry et al.(2018)]{Henry2018} Henry A., Berg D. A., Scarlata C.,
Verhamme A., Erb D., 2018, \apj, 855, 96  

\bibitem[Huang et al.(2016)]{Huang2016} Huang K.-H.,  Brada\'{c} M.,
 Lemaux B. C. et al., 2016, \apj, 817, 11

\bibitem[Izotov et al.(1994)Izotov, Thuan \& Lipovetsky]{Izotov1994}
Izotov Y. I., Thuan T. X., Lipovetsky V. A., 1994, \apj, 435, 647

\bibitem[Izotov et al.(2004)]{Izotov2004a} Izotov Y. I., Stasi\'nska G.,
Guseva N. G., Thuan T. X., 2004, \aap, 415, 87 

\bibitem[Izotov \& Thuan(2004)]{IzTh2004} Izotov Y. I., Thuan T. X., 2004,
\apj, 602, 200  

\bibitem[Izotov et al.(2006a)]{IzStasMeynet2006a} Izotov Y. I., Stasi\'nska G.,
Meynet G., Guseva N. G., Thuan T. X., 2006a, \aap, 448, 955 

\bibitem[Izotov et al.(2006b)]{Izotov2006b} Izotov Y. I., Schaerer D., Blecha
A., Royer F., Guseva N. G., North P., 2006b, \aap, 459, 71 

\bibitem[Izotov et al.(2011a)]{Izotov2011} Izotov Y. I., Guseva N. G.,
  Fricke K. J., Henkel C., 2011a, \aap, 536, L7
  
\bibitem[Izotov et al.(2011b)Izotov, Guseva \& Thuan]{IzGusevaT2011}
Izotov Y. I., Guseva N. G., Thuan T. X., 2011b, \apj, 728, 161 

\bibitem[Izotov et al.(2012a)Izotov, Thuan \& Guseva]{IzTGuseva2012}
Izotov Y. I., Thuan T. X., Guseva N. G., 2012, \aap, 546, 122 

\bibitem[Izotov et al.(2012b)Izotov, Thuan \& Privon]{IzTPrivon2012}
Izotov Y. I., Thuan T. X., Privon G., 2012, \mnras, 427, 1229
  
\bibitem[Izotov et al.(2014a)Izotov, Thuan \& Guseva]{IzTGuseva2014}
Izotov Y. I., Thuan T. X., Guseva N. G., 2014a, \mnras, 445, 778

\bibitem[Izotov et al.(2014b)]{Izotov2014b} Izotov Y. I., Guseva N. G.,
  Fricke K. J., Henkel C., 2014b, \aap, 561, A33

\bibitem[Izotov et al.(2014c)]{Izotov2014c} Izotov Y. I., Guseva N. G.,
  Fricke K. J., Henkel C., 2014c, \aap, 570, A97 

\bibitem[Izotov et al.(2016a)]{I16a} Izotov Y. I., Orlitov\'a I., Schaerer D., 
Thuan T. X., Verhamme A., Guseva N. G., Worseck G., 2016a, \nat, 529, 178

\bibitem[Izotov et al.(2016b)]{I16b} Izotov Y. I., Schaerer D., Thuan, T. X., 
Worseck G., Guseva N. G., Orlitov\'a I., Verhamme A., 2016b, \mnras, 461, 3683

\bibitem[Izotov et al.(2017a)]{I17a} Izotov Y. I., Guseva N. G., Fricke K. J.,
  Henkel C. \& D. Schaerer, 2017a, \mnras, 467, 4118

\bibitem[Izotov et al.(2017b)Izotov, Thuan \& Guseva]{ITG2017diag}
  Izotov Y. I., Thuan T. X., Guseva N. G., 2017b, \mnras, 471, 548

\bibitem[Izotov et al.(2018a)]{I18a} Izotov Y. I., Thuan T. X., Guseva N. G.,
Liss S. E., 2018a, \mnras, 473, 1956

\bibitem[Izotov et al.(2018b)]{Izotov2018b} Izotov Y. I., Schaerer D.,
Worseck G., Guseva N. G., Thuan T. X., Verhamme A., Orlitov\'a I., Fricke K. J.,
2018b, \mnras, 474, 4514

\bibitem[Izotov et al.(2019)Izotov, Thuan \& Guseva]{Izotov2019} Izotov Y. I.,
Thuan T. X., Guseva N. G., 2019, \mnras, 483, 549 

\bibitem[Izotov et al.(2020)]{Izotov2020} Izotov Y. I., Schaerer D.,
Worseck G.,  Verhamme A., Guseva N. G., Thuan T. X.,  Orlitov\'a I.,
Fricke K. J., 2020, \mnras, 491, 468

\bibitem[James et al.(2014)]{James2014} James B. L., Pettini M.,
  Christensen L. et al., 2014, \mnras, 440, 1794

\bibitem[Jaskot \& Oey(2013)]{JaskotOey2013} Jaskot A. E., Oey M. S.,
2013, \apj, 766, 91

\bibitem[Kakiichi \& Gronke(2019)]{Kakiichi2019} Kakiichi K., Gronke M., 2019,
  arXiv:1905.02480 

\bibitem[Katz et al.(2020)]{Katz2020} Katz H., \^Durov\^c\'ikov\'a D., Kimm T.
  et al., 2020, arXiv:2005.01734

\bibitem[Kauffmann et al.(2003)]{Kauffmann2003} Kauffmann G., Heckman T. M.,
White S. D. M. et al., 2003, \mnras, 341, 33

\bibitem[Kehrig et al.(2015)]{Kehrig2015} Kehrig C., V\'ilchez J. M.,
  P\'erez-Montero E., Iglesias-Páramo J., Brinchmann J., Kunth D., Durret F., 
Bayo F. M., 2015, \apj, 801, L28 
  
\bibitem[Kehrig et al.(2018)]{Kehrig2018} Kehrig C., V\'ilchez J. M.,
Guerrero M. A., Iglesias-P\'aramo J., Hunt L. K., Duarte-Puertas S., 
Ramos-Larios G., 2018, \mnras, 480, 1081 

\bibitem[Kennicutt(1998)]{Kennicutt1998} Kennicutt R. C., Jr.,
1998, \araa, 36, 189

\bibitem[K\"oppen \& Hensler(2005)]{Koppen2005} K\"oppen J., Hensler G.,
  2005, \aap, 434, 531

\bibitem[Kewley et al.(2013)]{Kewley2013a} Kewley L. J., Dopita M. A.,
  Leitherer C., Dav\'e R., Yuan T., Allen M., Groves B., Sutherland R., 
2013, \apj, 774, 100

\bibitem[Kimm et al.(2019)]{Kimm2019} Kimm T., Blaizot J., Garel T., 
Michel-Dansac L., Katz H., Rosdahl J., Verhamme A., Haehnelt M.,
2019, \mnras, 486, 2215  

\bibitem[Kobulnicky \& Kewley(2004)]{KobulnickyKewley2004} Kobulnicky H. A.,
  Kewley L. J., 2004, \aj, 617, 240 

\bibitem[Kojima et al.(2017)]{Kojima2017} Kojima T., Ouchi M., Nakajima K., 
Shibuya T., Harikane Y., Ono Y., 2017, \pasj, 69, 44

\bibitem[Kriek et al.(2015)]{Kriek2015} Kriek M., Shapley A. E., Reddy N. A.
et al., 2015, \apjs, 218, 15

\bibitem[Loaiza-Agudelo et al.(2020)Loaiza-Agudelo, Overzier \& Heckman]{Loaiza2020} 
Loaiza-Agudelo M., Overzier R. A., Heckman T., 2020, \apj, 891, 19

\bibitem[Lodders(2010)]{Lodders2010} Lodders K., 2010, in Principles and
Perspectives 
in Cosmochemistry, Astrophysics and Space Science Proceedings,
ISBN 978-3-642-10351-3. Springer-Verlag Berlin Heidelberg, 379

\bibitem[Lodders(2020)]{Lodders2020} Lodders K., 2020, Solar Elemental
Abundances, in The Oxford Research Encyclopedia of Planetary Science, Oxford
University Press

\bibitem[Ma et al.(2020)]{Ma2020} Ma X., Quataert E., Wetzel A., Hopkins P. F.,
Faucher-Gigu\'re C.-A., Kere\v{s} D., 2020, arXiv:2003.05945 

\bibitem[Maiolino et al.(2015)]{Maiolino2015} Maiolino R., Carniani S.,
Fontana A. et al., 2015, \mnras, 452, 54

\bibitem[Maseda et al.(2020)]{Maseda2020} Maseda M. V., Bacon R., Lam D.
  et al., 2020, astro-ph:2002.11117

  \bibitem[Masters et al.(2014)]{Masters2014} Masters D., McCarthy P.,
 Siana B. et al., 2014,  \apj, 785, 153 

\bibitem[Moll\'a et al.(2006)]{Molla2006} Moll\'a M., V\'ilchez J. M.,
Gavil\'an M., D\'az A. I., 2006, \mnras, 372, 1069 

\bibitem[Naidu et al.(2018)]{Naidu2018} Naidu R. P., Forrest B., Oesch P. A.,
  Tran K.-V. H., Holden B. P., 2018, \mnras, 478, 791

\bibitem[Naidu et al.(2020)]{Naidu2019} Naidu R. P., Tacchella S., Mason C. A.,
  Bose S., Oesch P. A., Conroy C., 2020, \apj, 892, 109

\bibitem[Nakajima et al.(2013)]{Nakajima2013} Nakajima K., Ouchi M.,
 Shimasaku K., Hashimoto T., Ono Y., Lee J. C., 2013, \apj, 769, 3

\bibitem[Nakajima \& Ouchi(2014)]{Nakajima2014} Nakajima K., Ouchi M., 2014, 
\mnras, 442, 900

\bibitem[Nakajima et al.(2020)]{Nakajima2020} Nakajima K., Ellis R. S.,
Robertson B. E., Tang M., Stark D. P., 2020, \apj, 889, 161  

\bibitem[Nanayakkara et al.(2020)]{Nanayakkara2020} Nanayakkara T.,
  Brinchmann J., Glazebrook K. et al., 2020, \apj, 889, 180

\bibitem[Oesch et al.(2010)]{Oesch2010} Oesch P. A., Bouwens R. J.,
Carollo C. M. et al., 2010, \apjl, 709, L21

\bibitem[Ono et al.(2012)]{Ono2012} Ono Y., Ouchi M., Mobasher B. et al.,
2012, \apj, 744, 83

\bibitem[Ota et al.(2014)]{Ota2014} Ota K., Walter F., Ohta K. et al.,
2014, \apj, 792, 34

\bibitem[Overzier et al.(2009)]{Overzier2009} Overzier R. A., Heckman T. M.,
  Tremonti C. et al., 2009, \apj, 706, 203

\bibitem[Ouchi et al.(2013)]{Ouchi2013} Ouchi M., Ellis R., Ono Y. et al.,
 2013, \apj, 778, 102

\bibitem[Paulino-Afonso et al.(2018)Paulino-Afonso, Sobral \& Ribeiro]{Paulino-Afonso2018}
    Paulino-Afonso A., Sobral D., Ribeiro B., 2018, \mnras, 476, 5479

\bibitem[P\'erez-Montero et al.(2019) P\'erez-Montero,
  Garc\'ia-Benito \& V\'ilchez]{Perez2019} P\'erez-Montero E.,
  Garc\'ia-Benito R., V\'ilchez J. M., 2019, \mnras, 483, 3322

\bibitem[Pilyugin et al.(2012) Pilyugin, Grebel \& Mattsson]{Pilyugin2012}
Pilyugin L. S., Grebel  E. K., Mattsson L., 2012, \mnras, 424, 2316

\bibitem[Plat et al.(2019)]{Plat2019} Plat A., Charlot S., Bruzual G., 
Feltre A., Vidal-García A., Morisset C., Chevallard J., Todt H.,
  2019, \mnras, 490, 978

\bibitem[Ramambason et al.(2020)]{Ramambason2020} Ramambason L., Schaerer D.,
Stasi\'nska G., Izotov Y. I., Guseva N. G., V\'ilchez J. M., Amor\'in R., 
2020, \aap, in preparation  

\bibitem[Roberts-Borsani et al.(2016)]{Roberts-Borsani2016}
  Roberts-Borsani G. W., Bouwens R. J., Oesch P. A. et al., 2016,
  \apj, 823, 143

\bibitem[Robertson et al.(2013)]{Robertson2013} Robertson B. E.,
Furlanetto S. R., Schneider E. et al., 2013, \apj, 768, 71

\bibitem[Roy et al.(1992)]{Roy1992} Roy J.-R., Aube M., McCall M. L.,
  Dufour R. J., 1992, \apj, 386, 498

\bibitem[Salmon et al.(2015)]{Salmon2015} Salmon B., Papovich C.,
 Finkelstein S. L. et al., 2015, \apj, 799, 183

\bibitem[S\'anchez-Almeida et al.(2016)]{Sanchez-Almeida2016}
  S\'anchez-Almeida J., P\'erez-Montero E., Morales-Luis A. B. et al.,
  \apj, 819, 110

\bibitem[Sanders et al.(2016a)]{Sanders2016} Sanders R. L., Shapley A. E.,
  Kriek M. et al., 2016a, \apj, 816, 23

\bibitem[Sanders et al.(2016b)]{Sand2016} Sanders R. L., Shapley A. E.,
  Kriek M. et al., 2016b, \apj, 825, L23
  
\bibitem[Sanders et al.(2020)]{Sanders2020} Sanders R. L., Shapley A. E.,
  Reddy N. A. et al., 2020, \mnras, 491, 1427

\bibitem[Santini et al.(2017)]{Santini2017} Santini P., Fontana A.,
    Castellano M. et al., 2017, \apj, 847, 76

\bibitem[Schaerer(1996)]{Schaerer1996} Schaerer D.,
1996, \apj, 467, L17 

\bibitem[Schaerer \& de Barros(2009)]{SchaererdeBarros2009} Schaerer D.,
de Barros S., 2009, \aap, 502, 423

\bibitem[Schaerer et al.(2015)]{Schaerer2015} Schaerer D., Boone F.,
  Zamojski M., Staguhn J., Dessauges-Zavadsky M.,
Finkelstein S., Combes F., 2015, \aap, 574, A19

\bibitem[Schaerer et al.(2016)]{Schaerer2016} Schaerer D., Izotov Y. I.,
  Verhamme A., Orlitov\'a I., Thuan T. X., Worseck G., Guseva N. G., 
2016, \aap, 591, L8

\bibitem[Schaerer et al.(2019)]{Schaerer2019} Schaerer D., Fragos T.,
  Izotov Y. I., 2019, \aap, 622, L10

\bibitem[Shapley et al.(2019)]{Shapley2019} Shapley A. E., Sanders R. L.,
  Shao P. et al., 2019, \apjl, 881, L35

\bibitem[Shibuya et al.(2015)Shibuya, Ouchi \& Harikane]{Shibuya2015}  Shibuya T., Ouchi M.,
    Harikane Y., 2015, \apjs, 219, 15

\bibitem[Shirazi \& Brinchmann(2012)]{Shirazi2012} Shirazi M., Brinchmann J.,
  2012, \mnras, 421, 1043

\bibitem[Smit et al.(2014)]{Smit2014} Smit R., Bouwens R. J., Labb\'{e} I.
  et al., 2014, \apj, 784, 58

\bibitem[Smit et al.(2015)]{Smit2015} Smit R., Bouwens R. J., Franx M.
  et al., 2015, \apj, 801, 122

\bibitem[Stark et al.(2013)]{Stark2013} Stark D. P., Auger M., Belokurov V.
  et al., 2013, \mnras, 436, 1040

\bibitem[Stark et al.(2013a)]{Stark2013a} Stark D. P., Schenker M. A.,
Ellis R. et al., 2013a, \apj, 763, 129

\bibitem[Stark(2016)]{Stark2016} Stark D. P., 2016, \araa, 54, 761

\bibitem[Stark et al.(2017)]{Stark2017} Stark D. P., Ellis R. S.,
Charlot S. et al., 2017, \mnras, 464, 469
  
\bibitem[Steidel et al.(2014)]{Steidel2014} Steidel C. C., Rudie G. C.,
  Strom A. L. et al., 2014, \apj, 795, 165

\bibitem[Steidel et al.(2016)]{Steidel2016} Steidel C. C., Strom A. L.,
  Pettini M., Rudie G. C. Reddy N. A., Trainor R. F., 2016, \apj, 826, 159  

\bibitem[Storey \& Hummer(1995)]{Storey1995}  Storey P. J., Hummer D. G.,
  1995, \mnras, 272, 41 

\bibitem[Strom et al.(2017)]{Strom2017} Strom A. L., Steidel C. S., Rudie G. C.,
Trainor R. F., Pettini M., Reddy N. A., 2017, \apj, 836, 164

\bibitem[Strom et al.(2018)]{Strom2018} Strom A. L., Steidel C. S., Rudie G. C.
Trainor R. F., Pettini M., 2018, \apj, 868, 117

\bibitem[Szecsi et al.(2015)]{Szecsi2015} Szecsi D.,  Langer N., Yoon S.-C.,
Sanyal D., de Mink S., Evans C. J., Dermine T., 2015, \aap, 581, 15 

\bibitem[Tang et al.(2019)]{Tang2019} Tang M., Stark D., Chevallard J.,
Charlot S., 2019, \mnras, 489, 2572 

\bibitem[Tayal(2007)]{Tayal2007} Tayal S. S., 2007, \apjs, 171, 331

\bibitem[Tayal \& Zatsarinny(2010)]{TayalZat2010} Tayal S. S., Zatsarinny O.,
  2010, \apjs, 188, 32

\bibitem[Thuan \& Izotov(2005)]{TI2005} Thuan T. X., Izotov Y. I., 2005, \apj,
\apjs, 161, 240

\bibitem[Topping et al.(2019)]{Topping2019} Topping M. W., Shapley A. E.,
Reddy N. A., Sanders R. L., Coil A. L., Kriek M., Mobasher B. Siana B., 
2019, MNRAS, arXiv:1912.10243 

\bibitem[Vincenzo et al.(2016)]{Vincenzo2016} Vincenzo F., Belfiore F.,
  Maiolino R., Matteucci F., Ventura P., 2016, \mnras, 458, 3466 

\bibitem[Watson et al.(2015)]{Watson2015} Watson D., Christensen L.,
Knudsen K. K., Richard J., Gallazzi A.,
Michałowski M. J., 2015, \nat, 519, 327

\bibitem[Yang et al.(2017)]{Yang2017} Yang H., Malhotra S., Gronke M. et al.,
  2017, \apj, 844, 171

\end{thebibliography}
\end{document}